\documentstyle[preprint,aps,eqsecnum]{revtex}
\newcommand{\prL}{Phys.\ Rev.\ Lett.\ }
\newcommand{\pr}{Phys.\ Rev.\ }
\newcommand{\jpb}{J.\ Phys.\ B: Atom.\ Mol.\ Opt.\ Phys.}

\newcommand{\phr}{Phys.\ Rep.\ }

\newcommand{\anp}{Ann.\ Phys.\ }

\newcommand{\csf}{Chaos, Solitons and Fractals\ }

\newcommand{\jpa}{J.\ Phys.\ A\ }

\def\bea{\begin{eqnarray}}
\def\eea{\end{eqnarray}}
\def\be{\begin{equation}}
\def\ee{\end{equation}}
\def\c#1{\setbox0=\hbox{#1}\ifdim\ht0=1ex\accent24 #1%
  \else{\ooalign{\hidewidth\char24\hidewidth\crcr\unhbox0}}\fi}
\begin{document}
\tighten
\author{Krzysztof Sacha$^{1,2}$, 
Jakub Zakrzewski$^2$, and Dominique Delande$^3$  }
\address{
$^1$ Fachbereich Physik,
Univerisit\"at Marburg,\\
Renthof 6,
D-35032 Marburg,
Germany\\
 $^2$Instytut Fizyki imienia Mariana Smoluchowskiego, Uniwersytet
Jagiello\'nski,\\
ulica Reymonta 4, PL-30-059 Krak\'ow, Poland\\
$^3$Laboratoire Kastler-Brossel, Tour 12, \'Etage 1, Universit\'e Pierre
et
Marie Curie,\\
4 Place Jussieu, 75005 Paris, France
}
\title{Breaking time reversal symmetry in chaotic driven Rydberg atoms
}
\date{\today}

\maketitle

\begin{abstract}
We consider the dynamics of Rydberg states of the hydrogen atom driven
by
a microwave field of elliptical polarization, with 
a possible additional static electric field. 
We concentrate on the effect of a resonant weak field - whose frequency
is close to the Kepler frequency of the electron around the nucleus -
which essentially produces no ionization of the atom, but completely
mixes the various states inside an hydrogenic manifold of fixed
principal quantum number. For sufficiently small fields, a perturbative
approach (both in classical and quantum mechanics) is relevant.
For some configurations
of the fields, the classical secular motion 
(i.e. evolution in time of the elliptical 
electronic trajectory) is shown to be predominantly chaotic. Changing 
the orientation of the static field with respect to the polarization
of the microwave field allows us to investigate the effect of
generalized time-reversal symmetry breaking on the statistical
properties
of energy levels.
\end{abstract}
\draft
\narrowtext
\newpage

\section{INTRODUCTION}

Sixteen years ago, Bohigas, Giannoni and Schmit, in a seminal paper
\cite{bohigas84},
formulated the conjecture that quantum systems which are 
chaotic in the classical limit, generically have statistical properties
of
energy levels described by Random Matrix Theory (RMT)
\cite{mehta91}.
While RMT had proven useful in characterizing nuclear spectra
\cite{brody81},
the conjecture was quite surprising since it claimed an applicability of
RMT statistics to deterministic systems with very few degrees of
freedom. 
While some counterexamples exist (see e.g. \cite{kuba95}),
the conjecture remains widely
accepted and played a very stimulating role in quantum chaos studies
(see for reviews \cite{haake90,bohigas91}).

Depending on the symmetries of a given strongly chaotic system, the
statistical
properties of its quantum spectrum fall into one of the three classes
known 
from RMT: orthogonal, unitary and symplectic. The orthogonal class
is associated with Hamiltonians invariant with respect to 
some anti-unitary symmetry. 
Typically this is true for time-reversal invariant systems,
but also for more complicated anti-unitary symmetry 
(generalized time-reversal symmetry) as the product of a time-reversal
with some
discrete reflection. Roughly speaking, this happens 
if a basis can be chosen where the matrix elements of the Hamiltonian 
are all real.
The corresponding statistical ensemble of (real symmetric) 
random matrices is referred to as 
the Gaussian Orthogonal Ensemble (GOE).
In the absence of any anti-unitary symmetry, the corresponding class of
complex hermitian matrices is known
as the Gaussian Unitary Ensemble (GUE). 
Finally, specific considerations apply to half-integer spin systems
with (generalized) time-reversal symmetry. Indeed, the energy levels
of these systems are systematically two-fold degenerate (this is the
well-known Kramers degeneracy). In the presence of a geometrical
symmetry
like azimuthal symmetry, the Kramers degeneracy is hidden and the GOE
statistics apply. With no geometrical symmetries, the  
Gaussian Symplectic Ensemble (GSE) has to be used. 
All these three classes of systems are characterized by level repulsion,
that is
zero probability of observing accidentally degenerate energy levels.
This is to be contrasted with a generic behaviour of multidimensional 
integrable systems where close lying levels are uncorrelated and obey
a Poisson statistics \cite{berry77}.
For a generic Hamiltonian system, where typically
chaotic and regular motions coexist, the statistical properties
of energy levels are not universal and are expected to be
intermediate between the two limiting cases, Poisson and RMT.

While the conjecture has been tested on a number of theoretical models
(we refer the reader to reviews \cite{haake90,bohigas91} rather than
numerous original papers), experimental verifications are much less
abundant and restricted fully, as far as we know,
to the orthogonal universality class \cite{zimmermann88}. This
is probably due to the fact that most experimental results are
obtained for atomic and molecular spectra. In the presence of a 
uniform magnetic field, the time-reversal symmetry is broken, but
a anti-unitary symmetry persists\cite{haake90}. Thus, observing GUE
statistics should require well controlled inhomogeneous fields on the
atomic scale, which has not been achieved.

An alternative is to study the eigenmodes not of the Schr\"odinger
equation, but of a different although similar wave equation. 
The best examples are two-dimensional microwave billiards (in fact
three-dimensional flat cavities below the cut-off frequency)
where the classical  Helmholtz
wave equation is in fact equivalent to the time-independent
Schr\"odinger equation.
There, breaking of the time reversal invariance is possible
by using magnetic devices and 
GUE-type statistics have been observed experimentally
\cite{so95,stoffregen95}. Still 
it is  desirable to have at hand 
a quantum system where manifestations of the
generalized time reversal symmetry breaking is experimentally
accessible.
The aim of this paper is to discuss an example of such a system. 
As mentioned above, breaking all anti-unitary symmetries using static
fields
is not possible. Hence, the idea is to use a time-dependent field acting
on an atom. With an oscillatory field alone, the product of time
reversal by
symmetry with respect to one of the polarization axis is an anti-unitary
symmetry. Hence, GUE statistics requires to combine a microwave field
and a static field.

The model we use, the Rydberg states of the hydrogen atom driven 
by a microwave field, has a 
long history of its own.
The system
attracted attention when it turned out that experimental 
results \cite{bayfield74} on the ionization probability
as a function of the microwave field and frequency
could be explained in terms of the underlying classical chaotic dynamics
which results in a diffusive energy gain of the Rydberg electron
and eventually to ionization\cite{leopold78}. 
Many interesting  phenomena (e.g. quantum localization) have been
studied
on this model both theoretically and experimentally (see e.g. review
papers 
\cite{casati87c,casati88,koch92b,koch95b}). The first experiments
involved 
linearly polarized microwaves allowing for a rough description of the
ionization thresholds via a 
simple one-dimensional time-dependent model where the motion
of the electron is restricted to the microwave field
axis\cite{jensen91}. 
 For other polarizations (available in recent experiments
 \cite{bellermann96,bellermann97}), such a simple model is not possible:
 the simplest dynamics may be two-dimensional
where the electronic motion is restricted to the polarization plane.
Such models have been investigated both for circular polarization (CP) 
\cite{kuba96,delande97b,sacha97c} and general elliptical polarization
(EP)
\cite{sacha97,sacha98b,sacha99a}.

When not only the ionization thresholds, but also the more subtle
details of the dynamics, have to be studied, a full three dimensional
atom
must be considered.
For LP microwave, the conservation of
the angular momentum projection onto the polarization axis, $L_z$, makes
the dynamics effectively two-dimensional and time dependent. For CP
case,
while $L_z$ is not conserved, the transformation to the frame rotating
with the microwave frequency removes the explicit oscillatory
time-dependence, leading to a three-dimensional
time independent problem. Both these simplifications are no longer
possible in the general EP microwave field and  the problem
becomes truly three dimensional and time dependent, 
providing new challenges to the theory.
It is this situation which we shall consider later on.

However, instead of discussing typical ``large'' microwave amplitude
for
fields which lead to an efficient excitation of Rydberg atoms to higher
excited states and finally to ionization, we shall
consider a ``weak'' perturbation of an atom where the exchange of energy
between the field and the atom is negligible.
The fact that the Coulomb problem is 
highly degenerate provides us with the possibility, as shown below,
to reach chaotic dynamics even for infinitesimally weak external fields.
This is not in contradiction with the Kolmogorov-Arnold-Moser (KAM)
theorem
\cite{lichtenberg83}
since the latter precisely does not apply for degenerate situations.
Those weak fields do not really excite the atom, but rather couple and
mix among each other the $n_0^2$ states of a degenerate hydrogenic
manifold 
with a given principal quantum number $n_0$. 
The degeneracy is lifted, states undergo
small (but measurable) shifts that reflect the dynamics of the $n_0$
manifold.
Such a situation is referred to as intramanifold dynamics (by contrast
to
a typical intermanifold coupling leading to atomic excitation and
ionization).
An intramanifold chaotic behaviour is quite attractive from the
theoretical
point of view. It yields an effective quantum Hamiltonian acting in the
$n_0$ space,
whose eigenvalues are the energy shifts, and is represented by a 
{\it finite} dimensional matrix of size  
$n_0^2$ (modulo the remaining point symmetries) with
no cutoff errors. The semiclassical limit is realized by letting
$n_0\rightarrow \infty$.

The first attempt to produce some intramanifold chaos used
the hydrogen atom in uniform crossed magnetic and electric fields 
\cite{uzer94,uzer97}, which was later extended to arbitrary mutual
orientations of the two fields \cite{main98}.
The authors, exploring second order perturbation theory, have
observed signatures of intramanifold chaos in the quantum spectrum. 
However, the situation is somewhat complex: the first order term (in the
two external fields) is always integrable; it is only when combined
with a second order term of a comparable magnitude that some
noticeable region of chaos can be created. But this implies the
application of
large fields, and higher order terms in the
perturbative expansion are of importance. 
That, unfortunately, has been neglected by the authors.
Moreover, the high electric field values used lead then to an extremely
fast field ionization which blurs any long-time effects in the dynamics.
Especially, there are no more bound states in these conditions,
but only broad overlapping resonances, which implies that the 
statistical properties cannot be described by one of the three standard 
ensembles of random matrices \cite{remark1}.

Recently, we have proposed \cite{sacha99b}
another system revealing chaotic intramanifold dynamics
which does not suffer from the mentioned deficiency.
This is an hydrogen atom driven resonantly (i.e. with the frequency
$\omega$
which is an integer multiple of the Kepler frequency $\omega_{\mathrm
K}=1/n_0^3$)
by an elliptically polarized microwave with
the possible addition of some static weak electric field. The latter
allows us to 
break any generalized time-reversal symmetry of the system.

Since the driving field is periodic, by applying the Floquet theorem,
one can find the eigenstates of the system 
(the so-called Floquet or dressed states). 
The eigenenergies of the system are referred to as quasienergies of the
system 
and are defined modulo $\hbar\omega$. As we have discussed shortly for a
$2:1$
resonance \cite{sacha99b}, the statistical properties of the
quasienergies
reveal convincingly the symmetry breaking. 
Here we discuss the same system, i.e. 
an hydrogen atom driven by 
a resonant EP microwave field, both in the presence and in the absence
of the static
field. 

The paper is organized as follows. In Sec.~II we derive 
the quantum perturbative Hamiltonian for our system 
and its semiclassical counterpart.
They are  used in Sec.~III to analyze the behavior of the atom 
in a pure microwave field and in Sec.~IV for a
microwave field combined with a static electric field. The summary 
and the future perspectives form the content of the concluding section.

\section{PERTURBATION APPROACH }

We consider a realistic three-dimensional hydrogen atom placed in a
static
electric field and driven by an elliptically polarized microwave field. 
We define the $z$-axis as perpendicular to the plane of the polarization
of the
microwave field. The Hamiltonian
of the system, in atomic units, neglecting relativistic effects,
assuming
infinite mass of the nucleus, and employing dipole approximation reads:
\be
H=\frac{\mathbf p^2}{2}-\frac{1}{r}+F(x\cos\omega t+\alpha y\sin\omega
t)+
{\mathbf E}\cdot {\mathbf r},
\label{h}
\ee
where $F$, $\alpha$ and $\omega$ stand, respectively,
for the amplitude, degree of elliptical polarization and frequency of 
the microwave field while ${\mathbf E}$ denotes the static electric
field.

As mentioned in the Introduction, we are interested in effects due to
weak external fields. In the absence of any external field, the energy
levels
of the system are $-1/2n_0^2$ with $n_0$ the principal quantum number;
the degeneracy
of the $n_0$ hydrogenic manifold is $n_0^2.$ Even
very weak fields mix strongly states within the manifold; on the other
hand,
it is perfectly justified to treat coupling to other manifolds
perturbatively.
We shall do so first quantum-mechanically with the help of the effective
Hamiltonian approach \cite{cct92}. Then, we construct its
(semi-)classical
equivalent. The latter allows us to study the character of the classical
motion
and search for parameters corresponding to chaotic dynamics.

\subsection{Quantum perturbation method}

For any time-periodic Hamiltonian, the Floquet theorem states
that the most general solution of the Schr\"odinger equation
can be written as a linear combination of the ``Floquet states",
which are time-periodic functions and eigenstates of the
so-called Floquet Hamiltonian ${\cal H}_{\mathrm Floquet}$
\cite{shirley65}
of the system:
\be
{\cal H}_{\mathrm Floquet} \mid \phi (t) \rangle =
\left(H-i\frac{\partial}{\partial t}
\right)\mid \phi(t) \rangle=\epsilon\mid \phi(t) \rangle
\label{floquet}
\ee
where $\epsilon$ and $\mid \phi(t) \rangle$ are 
respectively the quasienergy and the
Floquet state. 
The solutions of Eq.~(\ref{floquet}) have to satisfy the usual boundary
conditions in configuration space and periodic boundary condition
in the time coordinate (by construction, Floquet states are
time-periodic).
Hence, the Floquet Hamiltonian ${\cal H}_{\mathrm Floquet}$ acts on an
extended Hilbert space
containing also the time coordinate.

For studying the quasi-energy spectrum, is it convenient to choose,
as a basis of the atomic Hilbert space, the Sturmian functions 
(see e.g. \cite{abu94,abuth,sachath} for details of this application)
$\mid n,L,M\rangle^{(\Lambda)}$, where
$L$ and $M$ are the usual angular 
and magnetic quantum 
numbers
respectively while $n\geq L+1$ labels the radial functions whose number
of nodes is $n-L-1.$ $\Lambda$ is a scaling parameter (unit of length
in configuration space) for the Sturmian functions.
For $\Lambda=n_0,$ the Sturmian functions
with $n=n_0$ are the exact hydrogenic states, 
eigenstates of the unperturbed atom.
As we intend to describe the dynamics inside
the $n_0$-manifold of the hydrogen atom, we choose to keep $\Lambda=n_0$
in all
calculations.

Along the time coordinate, we choose the usual oscillating exponential
functions as a basis of time-periodic functions. They are labeled
with an integer index $K$ and defined by:
\be
\langle t | K \rangle = e^{-iK\omega t}
\ee

The whole Hilbert space of the atom + periodic perturbation system
(Floquet
Hamiltonian) is spanned by
the tensor product of the configuration space and time basis: 
\be
\mid n,L,M,K \rangle=\mid n,L,M\rangle^{(n_0)} \otimes |K \rangle .
\ee

In the dressed atom language, $K$ may be loosely
identified with the number of photons exchanged between the atom and the
field.

The Sturmian basis is not orthogonal, but it satisfies the following
relation:
\be
\sum_{n,L,M,K}{| n,L,M,K \rangle \langle n,L,M,K | \frac{1}{2r}} = 1
\label{unity}
\ee
The advantage is that, when written in the Sturmian basis, all
the matrix elements
representing the various parts of the Floquet Hamiltonian have
some strong selection rules. The selection rules on $K$ trivially come
from the Fourier expansion of the time dependences; the selection
rules on $L$ and $M$ originate from the angular dependence of the
various operators. The selection rules on $n$ are far from obvious
and are at the heart of the definition and properties of the
Sturmian functions \cite{remark2}.
 
If we define:
\bea
H_0 &=& \frac{\mathbf p^2}{2}-\frac{1}{r}-i\frac{\partial}{\partial t}, 
\cr
U &=& F(x\cos\omega t+\alpha y\sin\omega t),\cr
V &=& {\mathbf E}\cdot {\mathbf r},
\eea
we obtain the following selection rules for the matrices
representing these operators in the Sturmian basis\cite{delandeth}:
\be
\begin{array}{lllll}
\Delta n=0,\pm 1, &  \Delta L=0, & \Delta M=0, &  \Delta K=0, &
\mbox{for\ } H_0 \mbox{\ and\ unity\ operators}, \cr
\Delta n=0,\pm 1, \pm 2 &  \Delta L=\pm 1, & \Delta M=0, \pm 1 &  
\Delta K=0, & \mbox{for\ } V, \cr
\Delta n=0,\pm 1,\pm 2, & \Delta L=\pm 1, & \Delta M=\pm 1, & \Delta
K=\pm 1, & 
\mbox{for\ } U.
\label{selr}
\end{array}
\ee

In addition, all matrix elements are known in closed forms and
involve only square roots of rational functions of the various
quantum numbers. 
Note that, because of the non orthogonal character of the Sturmian
basis, calculating
the Floquet quasi-energies requires to solve a generalized eigenvalue
problem
rather than a standard one. This is the price to pay for getting sparse
matrices.

The exact calculation of the quasi-energies is possible only
numerically.
However, we are interested in the situation where both the static
and the microwave fields are weak. Thus, a perturbative expansion is
convenient.
We will now perform it at the lowest non-vanishing order
for each external field.
Because, the zeroth order eigenstates are highly degenerate, we have to
use
degenerate perturbation theory. A convenient formulation is to use
an effective Hamiltonian which, at any order of the calculation, has
the same spectrum as the initial Hamiltonian, but acts only inside
the degenerate hydrogenic manifold. The details of the method are given
in \cite{cct92}. 
At first order, the calculation is trivial and the effective
Hamiltonian
$H^{(1)}$ is just the projection of the perturbation onto the 
manifold we are considering.
If $P$ denotes the projector onto the
degenerate $(n_0,K=0)$ manifold, it is simply:
\be
H^{(1)} = P (U+V) P = PVP
\ee
since  $U$ always changes $K$ by one unit.
The non-zero matrix elements of $H^{(1)}$
are those of $V$ with $\Delta n =0.$ Thus $H^{(1)}$
 is proportional to the static electric field.

The lowest non-vanishing contribution of the microwave field is at
second
order. It has the following well-known formal expression:
\be
H^{(2)} = P U Q \frac{1}{E_0-H_0} Q U P
\ee
where $E_0=-1/2n_0^2$ is the unperturbed energy of the hydrogenic
manifold and $Q=1-P$ is the projector onto the subspace complementary
to the hydrogenic manifold. 

Explicit calculation of $H^{(2)}$ is not straightforward. Indeed, if one
expands the $1/(E_0-H_0)$ onto the eigenstates of $H_0,$ one obtains
a infinite sum over the discrete states and continuum of the atomic
spectrum. The trick is to use the non-orthogonal Sturmian basis defined
above. Indeed, the projector $P$ has the following simple expression
is this basis:
\be
P= \sum_{L,M}{| n_0,L,M,0 \rangle \langle n_0,L,M,0 | \frac{1}{2r}},
\ee
and consequently
\be
Q= \sum_{L,M,(n,K)\neq (n_0,0)}{| n,L,M,K \rangle \langle n,L,M,K | 
\frac{1}{2r}}. 
\ee

The last step is to calculate the matrix element of the $1/(E_0-H_0)$
operator in the Sturmian basis. It is simply accomplished by noting
that the operator $(E_0-H_0)$ is diagonal in $L$,$M$ and $K$ and
tridiagonal
in $n$ (i.e. connects only state $n$ to states $n-1$,$n$ and $n+1$).
Thus the matrix elements of $1/(E_0-H_0)$ are simply obtained
by solving a triadiagonal set of coupled equations in each 
$(L,M,K)$ subspace coupled to the initial state.

Finally the whole effective quantum Hamiltonian inside the
$n_0$-manifold reads
\be
H_{\mathrm eff}=H^{(1)}+H^{(2)}.
\label{qeff}
\ee
This Hamiltonian takes into account the {\it direct} coupling between
the 
levels due to the presence of the static electric field (the term
proportional
to $E$) and the {\it indirect} coupling through all levels of other
manifolds,
i.e. process of absorption and emission of microwave photons (the term
proportional to $F^2$) \cite{cct92}.
Because of the selection rules on the $U$ and $V$ operators and the
simple algebraic structure of the effective Hamiltonian, $H_{\mathrm
eff}$
itself has the following selections rules: $\Delta n=0, \Delta K=0$ (by
construction)
and $\Delta L = 0,\pm 1,\pm 2,$ $\Delta M = 0, \pm 1, \pm 2.$
The diagonalization of $H_{\mathrm eff}$, i.e. of the sparse banded
matrix of dimension $n_0^2$,
 by standard routines, yields quasienergies of the system.

 The method has been tested in limiting situations, e.g., for
 parallel weak static and linearly polarized fields where quasienergies
 resulting from the effective Hamiltonian could be compared with exact
 diagonalization values. The results for EP presented below are obtained
 for field amplitudes for which excellent agreement between exact and
 perturbative results have been found in the limiting cases. 

 \subsection{Semiclassical perturbation method}

The general prescription for the semiclassical quantization of 
a time-periodic system 
has been described in \cite{breuer91}.
Recently, we used a similar procedure for a LP microwave \cite{abu98a},
for a static electric field parallel to a LP microwave \cite{sacha98a}
and for a two-dimensional model atom in the EP microwave case
\cite{sacha98b,sacha99a}.

The method requires first to define a classical Hamilton function
for which one can use the usual semiclassical quantization rules. 
This is done by passing 
to the extended phase space, defining the momentum $p_t$ conjugate to 
the $t$ (time) variable. It yields the new classical
Hamiltonian, 
${\cal H}_{\mathrm Floquet}=H+p_t$ 
\cite{lichtenberg83}, being the classical analog of the 
quantum Floquet Hamiltonian, Eq.~(\ref{floquet}).

As the next step, we express the Hamiltonian
in action-angle variables of the unperturbed Coulomb problem 
\cite{leopold86,sacha98c}.
Due to its high symmetry, several choices are possible.
The standard solution is to consider the canonically conjugate
pairs $(J,\Theta)$, $(L,\Psi)$ and $(M,\Phi)$. $J$
is the principal action (corresponding to the principal quantum number),
i.e.
the total action along an unperturbed Kepler elliptical trajectory of
the
electron. It is simply 
related to the size of the
ellipse. The corresponding angle, $\Theta$, determines the position of
the
electron on its elliptical trajectory and depends linearly on time, 
$\Theta\sim\omega_K t$, for an unperturbed atom. $L$ is the angular
momentum, $\Psi$ the angle of rotation around the axis
defined by the angular momentum vector. Similarly, $M,\Phi$
denote the projection of the angular momentum on the laboratory
$z$-axis and the angle of rotation around that axis, respectively.
The shape of the ellipse is best described by its eccentricity
$e=\sqrt{1-L^2/J^2}$ while its orientation in the configuration space is
determined by the Euler angles $(\Phi, \arccos M/L,\Psi)$ as defined by 
Goldstein \cite{goldstein80}. 

Using these canonical coordinates, it is possible to write down 
the full Floquet Hamiltonian. We now specialize to the resonant case
where the microwave frequency is an integer multiple of the
Kepler frequency of the unperturbed electron, i.e.  
$\omega_0=\omega/\omega_K =m$.
The corresponding action is:
\be
J = n_0 = \omega_K^{-1/3} = \left( \frac{\omega}{m} \right)^{-1/3}.
\ee
$n_0$ is interpreted as the principal quantum number of the
quantum hydrogenic manifold where the resonance takes place.
  
In the absence of any external field, the variables 
$(J,L,\Psi,M,\Phi)$ are all constant while $\Theta$ evolves
linearly in time (see above). Hence, in the presence of weak
external fields, the motion along the $\Theta$ variable will be much
faster than along the other coordinates and the secular perturbation 
theory \cite{lichtenberg83} can be used: it averages over the
nonresonant
terms and yields the approximate resonant Hamiltonian of the form
\be
{\cal H}_{\mathrm
sec}=-\frac{1}{2n_0^2}-\frac{3m^2}{2n^{4}_{0}}\hat{J}^{2}+
 F\Gamma_{m}\cos(\hat\Theta-\delta)+E\gamma+\hat p_t
\label{EPr}
\ee
where  
\bea
\gamma &=&-\frac{3}{2}n_0^2\left[\cos\varphi\sin\theta
          \left(\cos\Phi\cos\Psi-
\frac{M}{L}\sin\Phi\sin\Psi\right)\right. \cr
&&  +\sin\varphi\sin\theta\left(\sin\Phi\cos\Psi+\frac{M}{L}
\cos\Phi\sin\Psi\right) \cr
&& \left. +\cos\theta\sqrt{1-\frac{M^2}{L^2}}\sin\Psi
\right]\sqrt{1-\frac{L^2}{n_0^2}}.
\label{g}
\eea
and
\be
\hat\Theta=m\Theta-\omega t,\ \hat J=\frac{J-n_0}{m},
\ \hat p_t=p_t+\frac{\omega J}{m},
\label{rot}
\ee 
The 
secular variables $\hat{\Theta}, \hat{J}, \hat{p_t}$ are slowly
varying variables obtained by substracting the unperturbed resonant
quantities.
$\hat{\Theta}$ represents the phase drift of the electron along the
elliptical
orbit and $\hat{J}$ the distance (in action) to the exact resonance.

The orientation of the static field with respect to the $z$-axis is 
determined by the usual spherical angles, $\varphi$ and $\theta$. The
expression
for $\gamma$ looks complicated, but it is actually nothing but the
component of
the average atomic dipole on the static field axis.
Similarly, $\Gamma_m(L,\Psi,M,\Phi;\alpha)$ and
$\delta(L,\Psi,M,\Phi;\alpha)$
just represent
the amplitude and the phase of the atomic dipole at the microwave
frequency.
They can be obtained simply from the Fourier components of the
electron position on an  unperturbed elliptical trajectory.
The explicit, rather complicated formulae for
 $\delta(L,\Psi,M,\Phi;\alpha)$ and
$\Gamma_m(L,\Psi,M,\Phi;\alpha)$ 
are given by Eq.~(2.15) and Eq.~(2.16) of \cite{sacha98c}, respectively
and are reproduced in Appendix A for the convenience of the reader.

The last stage is to quantize the system using the approximate
Hamiltonian, 
Eq.~(\ref{EPr}). As any explicit time dependence has disappeared in the
secular
Hamiltonian, the quantization of $\hat p_t$ is trivial.
Taking into account that eigenstates of the Floquet Hamiltonian have to
be time-periodic, this yields $\hat p_t=k \omega$    
(where $k$ is an 
integer number) \cite{breuer91,abu98a} which ensures the periodicity of
the quasienergy spectrum with a period $\omega$.

The radial motion, i.e. in the $(\hat J,\hat\Theta)$
effectively decouples from the secular motion of the elliptical
trajectory, 
i.e. in the $(L,\Psi,M,\Phi)$ space
\cite{leopold86,leopold87}. While considering the radial motion, the 
effective hamiltonian for the secular part is approximately constant
(for a detailed discussion as well as possible counterexamples
in some cases see \cite{leopold86}). In effect the quantization 
resembles in
 spirit  the Born-Oppenheimer 
approximation. One may first quantize the radial motion keeping 
the secular motion frozen. The radial motion exhibits a pendulum-like
dynamics 
whose quantum eigenvalues are given by the solutions of the Mathieu
equation
(see the similar treatment for one-dimensional systems 
\cite{henkel92,holthaus95,flatte96,sirko95}). 
Hence, we can define an effective Hamiltonian acting in a reduced
$(L,\psi,M,\phi)$ phase space just replacing  the $(\hat J,\hat \Theta)$ 
part of ${\cal H}_{\mathrm sec}$ by the quantized energy levels of the
pendulum. In this paper, we are interested in the ground state of the
pendulum, 
thus, the quantization of the radial motion yields the following
effective Hamiltonian 
for the secular motion:
\be
{\cal H}_{\mathrm eff}=-\frac{3m^2}{8n_0^4}a_0(q)+E \gamma,
\label{semi}
\ee
where the constant terms $-1/2n_0^2$ and $k\omega$ are omitted
(${\cal H}_{\mathrm eff}$ denotes the shift from the unperturbed energy
level of the
atom), and:
\be
q=\frac{4n_0^4F}{3m^2}\Gamma_{m}.
\label{semiq}
\ee
is a dimensionless parameter.
$a_0(q)$ is the Mathieu parameter \cite{abrammowitz72}
corresponding to the ground state of the 
pendulum. We can introduce scaled quantities:
\bea
F_0=n_0^4F\\
E_0=n_0^4E\\
L_0=\frac{L}{n_0}\\
M_0=\frac{M}{n_0}\\
\Gamma_{m,0}=\frac{\Gamma_m}{n_0^2}\\
\gamma_0=\frac{\gamma}{n_0^2}.
\eea
The effective Hamiltonian is:
\be
{\cal H}_{\mathrm
eff}=-\frac{3m^2}{8n_0^4}a_0(q)+\frac{E_0}{n_0^2}\gamma_0.
\label{semiscaled}
\ee

For a large microwave field amplitude or in the deep semiclassical
limit, i.e.
$n_0\rightarrow\infty$, we may employ the asymptotic expression, for
large $q$,
of the Mathieu parameter \cite{abrammowitz72} 
\be
a_0(q)\rightarrow -2q+2\sqrt{q}.
\label{highf}
\ee 

This corresponds to the case where the pendulum is localized near its
stable equilibrium position (whose energy is $-2q$), its zero-point
energy in the ground state being calculated in the harmonic
approximation 
(hence the $2\sqrt{q}$ term). 

In the opposite limit, i.e. for $q\ll 1$, another
approximation exists $a_0(q)\approx-q^2/2$ \cite{abrammowitz72}.
This corresponds basically to a very weak trapping pendulum potential,
where
the motion is basically the free motion only slightly perturbed (at
second
order in the energy) by the potential. 
This is the classical counterpart of the quantum perturbative
approach developed above, the equivalent of the ``no $n$-mixing"
approximation.
In the following, we restrict
ourselves to this case as more appropriate for moderate $n_0$ and very
small $F_0$. 
Hence, the final expression for the effective secular 
Hamiltonian we are going to deal with is
\be
{\cal H}_{\mathrm eff}=\frac{F_0^2}{3m^2}\Gamma^2_{m,0}
+\frac{E_0}{n_0^2}\gamma_0.
\label{final}
\ee
This Hamiltonian is a semiclassical counterpart of the quantum effective
Hamiltonian, Eq.~(\ref{qeff}), namely first order in the static electric
field and second order in the resonant microwave field.

To calculate the quasienergies semiclassically, one should quantize 
the secular motion determined by the Hamiltonian (\ref{final}). 
It has been done in simpler situations (e.g. an hydrogen atom perturbed
by 
a linearly polarized microwave field and a
parallel
static electric field \cite{sacha98a}).
Then, the secular motion is integrable and its
quantization straightforward using the WKB quantization rule. 
The present secular problem has two degrees of
freedom and turns out to be non-integrable except for some limiting
cases.
Therefore, a detailed analysis of the classical motion in
the phase space of secular variables is necessary for possible
comparison with
quantal data.

\section{PURE MICROWAVE PERTURBATION CASE}

Let us consider first the
 perturbation of an hydrogen atom by an elliptically 
polarized microwave field in the absence of the static field.
 Our previous studies of resonant driving of the atom
were restricted to the simplified two-dimensional model atom where the
electronic motion is restricted to the polarization plane
\cite{sacha98b,sacha99a}.
Then, the classical secular motion is one-dimensional and application of
the WKB
quantization rule gives very accurate results for quasienergies of the
system. 
In a realistic three-dimensional model, a similar procedure is no longer 
possible as the effective classical Hamiltonian for a general
EP field is not integrable. 
The secular motion of the atom for weak microwaves, see
Eq.~(\ref{final}),
is determined by the Hamiltonian
\be
{\cal H}_{\mathrm
eff}=\frac{F_0^2}{3m^2}\Gamma^2_{m,0}(L_0,\Psi,M_0,\Phi;\alpha).
\label{purem}
\ee
The integrable motion is obtained in the
limiting polarization cases, i.e. $\alpha=0$ or $1$, only.
That is, for the LP field, the angular momentum projection on the 
polarization axis 
is a constant of motion. For the CP case, because of circular symmetry
of
the field, $\Phi$ becomes a cyclic variable and the secular motion is
also
effectively one-dimensional. Clearly, for $\alpha$ close to these
limiting values,
the secular motion will be close to integrable. With this in  mind, 
looking for chaotic dynamics, we take  $\alpha=0.4$ in the following 
(we have verified that this value is representative for a typical EP
behaviour). 

Eq.~(\ref{purem}) shows that 
the structure of the classical phase space of the secular motion depends
only 
on the value of ${\cal H}_{\mathrm eff}/F_0^2$ (beside the integer $m$
labeling the primary resonance). 
In other words, if the secular motion is non-integrable for some finite
field amplitude, it remains non-integrable even for infinitesimally
small
amplitude. This clearly demonstrates the inapplicability of the
Kolmogorov-Arnold-Moser theorem to the highly degenerate Coulomb
problem. 
On the other hand, the time scale of precession of an electronic ellipse
is
affected by the strength of the perturbation; for very small $F_0$,
it will be
extremely slow, but the trajectories of the secular motion do not depend
on $F_0.$

Consider the principal 1:1 resonance case, i.e. $m=1$. 
To focus on the phase
space structure, we have plotted Poincar\'e surfaces of section (SOS)
for a few values of ${\cal H}_{\mathrm eff}/F_0^2$ in Fig.~\ref{one}. 
For high values of the secular
energy, the motion is generally regular. However, for an energy interval
around
${\cal H}_{\mathrm eff}/F_0^2=0.3,$ the mixed character of the motion is
apparent,
a quite
large chaotic layer is clearly visible.
 
Switching to the 2:1 resonance case (right hand side of Fig.~\ref{one}) 
we find, as previously, regular phase space structures for high 
energy and mixed character of the motion for lower energies. By visual
inspection, the secular motion for $2:1$ resonance looks ``more
chaotic''
with smaller regular islands embedded in a pronounced chaotic layer. 
Although we
have searched quite extensively, we could not find values of $\alpha$
and the energy corresponding to fully chaotic motion. Always at best 
tiny regular islands have been found.

The mixed character of the secular motion should have consequences on
the
statistical properties of the quasienergy spectrum inside the
$n_0$ manifold. As the system possesses anti-unitary symmetry
invariance, 
i.e. is invariant under
time-reversal combined with the $y\rightarrow -y$ transformation, 
see Eq.~(\ref{h}), the statistical properties are expected to reflect
intermediate behavior between the Poisson and GOE character. To
calculate the
quasienergy spectrum, we employ the quantum effective Hamiltonian
Eq.~(\ref{qeff}). One should take care of discrete symmetries of the 
system. That is, the system is invariant under the $z\rightarrow -z$ 
transformation as well as the parity combined with translation in time 
by $\pi/\omega$ transformations. Thus, the whole spectrum of the
$n_0$-manifold 
splits into four independent smaller spectra which are unfolded 
independently in order to study level statistics. 

The dynamics of the levels belonging to the $n_0=20$ manifold as a
function
of the polarization degree $\alpha,$ are
shown in Fig.~\ref{two}, for the 1:1 and 2:1 resonance cases. 
In each panel, for clarity, there is only one of the four independent 
sub-spectra plotted. Qualitatively, the level dynamics reflects the
character 
of classical motion, i.e. for high energies, one cannot see level
repulsion
and there are apparently level crossing (actually small avoided
crossings).
At lower energy, the level dynamics is more irregular with plenty of 
avoided crossings, a clear signature of classical chaos in the system,
compare Fig.~\ref{twobis} which shows this region in more detail.

To make the comparison more quantitative, we have calculated the
cumulative nearest
neighbor spacing (NNS) distributions, taking levels in the energy
intervals ${\cal H}_{\mathrm eff}/F_0^2 \in [0.02-0.045]$ and 
$[0.00035-0.0018]$  for the 1:1
and 2:1 resonance cases, respectively. 
The intervals have been chosen to correspond to 
chaotic behavior as much as possible. Fig.~\ref{three} presents the
results
for the 1:1 resonance for principal quantum number $n_0$ around 55 and
for about twice bigger $n_0$, i.e. around 100. The similar results 
corresponding to the 2:1 resonance are plotted in Fig.~\ref{four}. 
In all cases, one can observe that the numerical data are intermediate 
between the Poisson and GOE statistics. However, the behavior of the 1:1 
case is closer to Poisson while the 2:1 one to GOE character, in
agreement
with the more classically chaotic behaviour in the latter case. 

Quantitative measures can be obtained by fitting the data to some
theoretical NNS distributions. Quite often, this is being done 
directly on the histograms for the NNS distributions. Such a procedure
is, 
however, bin size
dependent and should be avoided. We prefer
to fit the integrated (cumulative) distributions which do not suffer
from this
ambiguity. There are several possible choices for the theoretical
distributions.
Berry-Robnik statistics \cite{berry84} corresponds to a superposition of 
{\it independent} Poisson and
GOE spectra with  
relative weight
 $q$ corresponding 
to the relative volumes of regular and chaotic parts
in the classical phase space. 
Others possibilities are 
the phenomenological Brody \cite{brody73} and Izrailev \cite{izrailev90} 
distributions. 
The explicit expressions for the distributions used can be found in
Appendix B. Although the Berry-Robnik distribution relies on some
reasonable theoretical grounds while the two other ones are
purely phenomenological, it is commonly accepted that,
for not very small effective $\hbar$, the Brody distribution
-- or much less known Izrailev one --  works better than the
Berry-Robnik
proposition. The latter works well only in the very deep semiclassical
limit
(very small effective $\hbar$) where tunneling between regular and
chaotic 
parts of the phase space is negligible. For lower lying states, the
``regular'' 
and ``irregular'' part of the spectrum interact via tunneling, leading
to level
repulsion between states in the two groups.

This behaviour is clearly visible in Figs.~\ref{three} and \ref{four},
(b) and (d).
There, the Berry-Robnik distribution predicts much more small
spacings than actually observed.
On the other hand, except for small spacings
we have found that Berry-Robnik statistics fits best in most of the
cases, 
compare
Fig.~\ref{three} and Fig.~\ref{four}.

The obtained
fitted values of the Berry-Robnik
parameter and the parameters for the Brody and Izrailev distributions
are 
given in Table~\ref{t1}. While, for the latter cases, the parameters
have 
little physical meaning, as mentioned above, $q$ in the 
Berry-Robnik distribution should measure the fraction of the chaotic
phase 
space. Qualitatively, the value
obtained agrees well with classical SOS plots.

We have also studied the
spectral rigidity, i.e. the $\Delta_3$ statistics
\cite{mehta91,brody81,bohigas91},
in order to obtain some information on the long range correlations in
the spectra
of our system.
Spectral rigidity gives an independent
information about the relative measure of the chaotic part of the
classical
phase space. For a superposition of independent Poisson and GOE spectra,
one obtains \cite{seligman85}
\be
\Delta_3=\Delta_3^{\it Poisson}((1-q)L)+\Delta_3^{\it GOE}(
qL).
\label{delta3}
\ee
We have fitted our numerical results with this distribution.
The results are plotted in Fig.~\ref{five} while values of 
the fitted parameter are put in Table~\ref{t1}.
It is well known that, at large $L$, the spectral rigidity
deviates from any universal behaviour and saturates. This is a
non-semiclassical effect which should take place for larger and larger
$L$ 
as the effective $\hbar $ goes to zero.
To fit the parameter $q$, we have
taken into account only data up to $L=10$ in order not to enter the
saturation
regions visible in Fig.~\ref{five}.

Comparing the Poincar\'e SOS, NNS distributions and $\Delta_3$
statistics of the calculated data,
the qualitative agreement between the classical
dynamics and the corresponding quantum statistical properties is
apparent. The fitted values for the relative measure of the chaotic part
of the 
phase space coming either from the NNS or the $\Delta_3$ distributions
agree perfectly and match well the visual aspect of 
the SOS.

The NNS distributions change a little when the
principal quantum number is modified.
The $n_0\approx 55$ case reveals slightly 
stronger level repulsion (and, therefore, a ``more
chaotic'' character) than the $n_0\approx 100$ case.
This suggests that some tiny regular structures in the classical phase
space are not resolved for $n_0=55$ but are for $n_0=100.$
The long range correlations
are more dramatically sensitive to $n_0$.
The saturation of $\Delta_3$ starts at about 
twice larger distance in $L$ for the higher $n_0$ value, see
Fig.~\ref{five}.
This is in agreement with the theory of Berry\cite{berry85} 
where the critical $L$ value scales as $1/\hbar$ 
(the effective Planck constant in our problem is
$1/n_0$). For large $L,$
the $\Delta_3$ statistics saturates at almost
twice bigger value for $n_0\approx 100$ than $n_0\approx 55$ again in
a qualitative agreement with theory. The latter \cite{berry85} predicts
($\Delta_3(\infty)\sim 1/\hbar$) for a regular spectrum and 
 ($\Delta_3(\infty)\sim \ln (1/\hbar)$) for a strongly chaotic system.
 We deal with an intermediate system with mixed phase space, thus
 we expect the numerical $\Delta_3(\infty)$ value to lie in between the
two
 limits. This is indeed the case.

Finally, let us briefly argue what happens for larger microwave field
amplitudes.
The secular motion, in the weak field limit, is determined by the
effective 
Hamiltonian, Eq.~(\ref{purem}), which in turn is a function of
$\Gamma_{m,0}$. 
Increasing the field amplitude one leaves the validity range of the
Eq.~(\ref{purem}) and enters the region where Eq.~(\ref{highf}) is
applicable.
Then the orbital electronic motion becomes localized inside a resonance
island.
The secular motion, however, remains unchanged because the new effective
Hamiltonian is again a function of $\Gamma_{m,0}$ only. 
This means that the spectral statistical properties for higher field
amplitude 
(of course not so big as to produce strong intermanifold mixing) 
will be the same as the ones presented here. For the high secular
energy the motion will be also regular. Thus, one may expect that
the nonspreading wavepackets predicted using 
the two-dimensional model \cite{sacha98b,sacha99a} will 
also exist in the real three-dimensional world. 

\section{MICROWAVE PLUS STATIC ELECTRIC FIELD}

In this section, we discuss the intramanifold behaviour of the
hydrogen atom exposed to a resonant microwave field and a static
electric field of arbitrary mutual orientation.

For small field amplitudes, the effective secular Hamiltonian is given
by
Eq.~(\ref{final}). The Hamiltonian is the sum of  two terms -- the first
one
proportional to $F_0^2$ (square of the scaled microwave field), 
the second one to $E_0$ (scaled static field). 
For arbitrary mutual orientations of the two fields, 
the two terms have incompatible symmetry properties and, when having
comparable
magnitudes, induce a globally chaotic behavior. 

Eq.~(\ref{final}) has some well defined scaling properties with the
field
strengths $F_0$ and $E_0$. Let us define the reduced microwave strength
\be
{\cal F}=F_0^2n_0^2/E_0=F^2n_0^6/E 
\ee
and the reduced Hamiltonian
\be
{\cal H}={\cal H}_{\mathrm eff}n_0^2/E_0=\frac{{\cal
F}}{3m^2}\Gamma^2_{m,0}
+\gamma_0.
\label{finscal}
\ee
The classical phase space structure depends only on the value of ${\cal
H}$ 
and ${\cal F}$ (beside the static field vector orientation and the
polarization
of the microwave field), 
but not on the detailed values of $n_0$, $E_0$, $F_0$ and the
secular energy. Of course, weaker fields imply a slower secular motion,
but
this does not affect the structure of the phase space. In the quantum
mechanical picture, the energy splitting of a degenerate hydrogenic
manifold
also depends on absolute values of the fields, but the structure of the
levels
does not.

The application of the static electric field allows us to break any
anti-unitary symmetry of the system. It is only for $E_x=0$ (i.e.
$\varphi=\pi/2$) or $E_y=0$ (i.e. $\varphi=0$) that the system has some
anti-unitary invariance, under the combination of the time-reversal
transformation with reflection with respect to the $yz$ or $xz$ plane,
 see  Eq.~(\ref{h}). 

In a previous letter \cite{sacha99b}, we have considered the system in
the 
case of the  2:1 resonance driving, i.e. when the microwave frequency is
twice
the Kepler frequency. 
Let us consider here first the principal 1:1
resonance, i.e. $m=1$.  A possible signature of the anti-unitary
symmetry
breaking would be to observe level repulsion with stronger than linear
repulsion
(i.e. with $P(s)\propto s^\beta$ with $\beta >1$ for small $s$). 
Clearly, it is desirable
 to have a predominantly chaotic classical dynamics, as a transition
from GOE to GUE statistics is expected. To this end, 
we have  to find values of
the fields parameters which maximize chaoticity of the system. After
rather extensive searches, we have found that $\alpha=0.4$, ${\cal
F}=10$ 
and $\theta=\pi/4$ are a good choice for that purpose. 
The remaining spherical angle $\varphi$ determining the
orientation of the static electric field vector, is used to control the
breaking
of the
anti-unitary symmetry. 

In Fig.~\ref{six}, we show Poincar\'e SOS
for a few values of ${\cal H}$, for two cases: $\varphi=0$ and
$\pi/2$. 
One can see that, for $\varphi\approx \pi/2$, a predominantly 
chaotic structure  
space exists in phase space for a large range of secular energies. 
For $\varphi=\pi/2,$ the generalized time reversal invariance holds, 
as mentioned above. Thus, to study the symmetry breaking, it is
interesting 
to e.g. decrease (or increase) $\varphi$ gradually, 
collecting quantum data for some 
$\varphi$ values. Note that the addition of the static field tends to
make the classical dynamics slightly  more chaotic, compare
Figs~\ref{one}
and \ref{six}.

The data are then analyzed as in the previous Section -- an example is 
shown in Fig.~\ref{seven} for two different
orientations of the electric field $\varphi=\{0.4\pi,\pi/2\}$. For 
each $\varphi$, data have been collected for principal quantum number 
$n_0$ in the range 50-59. 
Then, we have chosen levels from the scaled energy interval ${\cal H}\in
(0,1.4)$,
unfolded 
each spectrum and calculated NNS distributions and spectral rigidities.
The cumulative NNS distributions are plotted in Fig.~\ref{seven}.
One can see that the NNS 
distribution corresponding to the anti-unitary invariant case, 
$\varphi=\pi/2$, is close to, but does not reach completely the GOE
behavior. 
Similarly, for the
$\varphi=0.4\pi$ case, the distribution is very close, but does not
reach the GUE one.
Nevertheless, the symmetry breaking is apparent and the numerical
spectrum
at $\varphi=0.4\pi$ shows much more level repulsion than the GOE case,
which is a clear-cut signature of the breaking of any anti-unitary
symmetry.

To measure departures from the entirely ergodic behavior, we can also
use
the Berry-Robnik distribution. The Berry-Robnik model
for fully broken antiunitary symmetry
consists of the superposition of two independent Poisson and GUE
spectra. 
The results of the fits are collected in Table~\ref{t2}.  We have not
applied the Brody distribution to the broken anti-unitary invariance
case, as
the distribution is defined only for Brody parameter less than unity,
and does not make any sense in the unitary case.
The Izrailev distribution does not suffer this severe problem,
and contain all limiting cases ($\beta =0$ for Poisson statistics,
$\beta=1$ for GOE and $\beta=2$ for GUE).
We have thus fitted our results with this distribution too.
 
In all cases, the Izrailev distribution works much better than
the Berry-Robnik ansatz, presumably because chaotic motion occupies most
of
the phase space and regular regions are very tiny as seen from
Fig.~\ref{six}
and from the values of $q$ obtained.

The values of the fitted Berry-Robnik parameter are consistent with the 
character of the corresponding classical motion. An independent
information 
about the relative measure of a chaotic part of phase space comes from
the 
fit of the $\Delta_3$ statistics, Fig.~\ref{eight} [for broken
anti-unitary
invariance case $\Delta_3^{\it GOE}$ is substituted by $\Delta_3^{GUE}$
in
Eq.~(\ref{delta3})], which turns out to agree very nicely with the
values of
the Berry-Robnik parameter.

By gradually decreasing $\varphi,$ we may observe the {\it partial} 
symmetry breaking by studying the variation of fitted parameters with
$\varphi$. 
Such a transition has been analytically studied  for
gaussian random matrices \cite{pandey83,mehta83} where the two-point
correlation function was analytically found for an appropriate ensemble
which interpolated between the GOE and the GUE. 
A further link with the dynamics of
fully chaotic systems with partially broken antiunitary symmetry was
also established \cite{bohigas95}. These developments cannot be used
here since the dynamics in our case is not fully chaotic (as seen from
SOS plots and the non-integer level repulsion $\beta$ parameters for extreme
cases of preserved and broken antiunitary symmetry, see Table II).

On the other hand, the Izrailev distribution is quite suitable
 since it should be a reasonable approximation 
   both for a partial symmetry breaking and a mixed dynamics
  (it would be possible
to construct an analog of Berry-Robnik distribution for such a case but
it
would be of little practical importance). 

 Fig.~\ref{nine} summarizes the changes of the fitted 
Izrailev parameter $\beta$
(small $s$ repulsion) with $\varphi$. For $\varphi=\pi/2,$ it is minimal
and
equal to 0.85 (see also Table~\ref{t2}) for $n_0$ around 55.
With departure from the value $\varphi=\pi/2$ where the anti-unitary
symmetry exists,
it rapidly
increases  (filled circles) up to the maximal
value
1.73 for $\varphi=0.4\pi,0.6\pi$. For still lower (higher) values of
$\varphi,$
 the trend is
reversed and $\beta$ starts to decrease. This, at first glance, is a 
surprising effect (since the symmetry should not yet start to restore).
However, it is most probably due the fact that the classical motion
becomes more regular as $\varphi$ is far from $\pi/2$.
Observe in Fig.~\ref{six} that the SOS
around $\varphi=0$ is much more regular than for $\varphi=\pi/2$.

In order to test this hypothesis, we have used another $n_0$ value.
Indeed, if the dip of $\beta$ near $\varphi=\pi/2$ was due to classical
reasons,
it should not
depend on $\hbar$ ,i.e. $n_0$. Fig.~\ref{nine} shows also
the 
fitted $\beta$ parameters for $n_0$ around 100. Generally, 
the $\beta$ values obtained
are slightly larger than for lower $n_0$. Clearly, $\varphi$ starts to
decrease again around the value $\varphi=0.4\pi$. On the
other hand, the dip of $\beta$ when $\varphi$ goes to $\pi/2$
(where an anti-unitary symmetry exists)
is faster than for lower $n_0$.  This is in a full qualitative
agreement with RMT \cite{haake90,bohigas91}: for the GOE $\rightarrow$
GUE
transition the  parameter controlling the transition is proportional to 
$N^{-1/2}$ where $N$ is the matrix dimension. The size of our matrices 
scales as $n_0^2$, so the parameter controlling the breaking (i.e.
a deviation from $\pi/2$ value) should scale like $n_0^{-1}\propto
\hbar_{\mathrm eff}$. Such a behaviour is roughly observed in Fig.10.

Fig.~\ref{ten} shows the ``maximal'' repulsion case
obtained,
i.e. data for $n_0$ around 100 and $\varphi=0.4\pi$. 
The numerical data are
presented in the form of the histogram of spacings and compared 
with the Izrailev distribution 
(the fit has been performed, as usual, for the cumulative
distribution; the resulting $\beta=1.83$ value has been used to plot
the Izrailev distribution). The dash-dotted  and dashed curves correspond 
to Wigner GOE and GUE 
distributions, respectively.
 The fact that we observe level repulsion much stronger than
the GOE behaviour is a signature of anti-unitary symmetry breaking.

As mentioned before, our first results on the manifestations of 
symmetry breaking
in our system have been obtained for the microwave frequency being
twice the Kepler frequency, i.e. for the 2:1 resonant driving
\cite{sacha99b}.
This choice  was motivated by SOS plots in the absence of the electric
field - see Fig.~\ref{one} - showing smaller regular islands for higher
microwave frequency. However, as we have seen above, the presence of the
electric field makes the secular motion in the principal
 resonance island
chaotic enough and in fact we get stronger repulsion for the $1:1$
resonance 
than for the $2:1$ situation reported before \cite{sacha99b}. For
completeness,
we show in Table~\ref{t3} the fitted parameters obtained from
the numerical results reported in \cite{sacha99b} 
using either the spacing distribution or the spectral
rigidity $\Delta_3.$ As can be seen, the conclusions we obtain from
these results completely confirm the analysis of the 1:1 resonance.

As far as we know, the studied system constitutes the first
experimentally
realizable example of a quantum system with broken anti-unitary
symmetry. 
We have considered small, but finite, field amplitudes to stay well
within the applicability range of the perturbation theory. Nevertheless,
this
is experimentally feasible: for $n_0\approx 55$ and 
$F_0\approx 5\times 10^{-4},$ i.e. about $0.3$ V/cm,
 the mean level spacing is of the order of
MHz.
   
For stronger fields, our classical studies also suggest a similar
behavior.
It could even be that the breaking of the secular approximation makes
the system more chaotic and that the statistical properties will be
closer
to GUE. 
However, we are not able to show quantum numerical results as they
require full
quantum numerical treatment which is difficult with
the present computer resources and must be left for a future work.
 
\section{CONCLUSIONS}

We have considered an hydrogen atom perturbed by a resonant elliptically
polarized microwave field with 
or without an additional static electric field of different orientation,
in the limit of small field amplitudes.
Classically, such fields may produce chaotic dynamics in the secular
motion of the
electronic elliptical trajectory. In quantum mechanical language, states
coming
from a given  hydrogenic manifold may be mixed significantly
only with each other. Such a situation has been interpreted 
as a signature of an intramanifold chaos. 

For the pure microwave problem, we have studied two different
resonant driving cases, i.e. the 1:1 and 2:1 resonances between the
microwave 
field and the unperturbed 
Kepler motion. Quantizing the fast orbital electronic motion, one can
derive
an effective Hamiltonian describing the slow secular precession 
of the electronic elliptical trajectory. 
For a generic elliptical polarization, the effective Hamiltonian has two
degrees of
freedom and turns out to be non-integrable. By means of Poincar\'e
surfaces of
sections, we have
found a range of the secular energy where the phase space reveals mixed
character with a significant amount of chaotic layer. Switching to a
quantum
perturbation calculation, we have shown that the statistical properties
of the
corresponding quasienergy levels reveal an intermediate behavior between
Poisson and GOE character. For the 2:1 resonance case, the classical
phase 
space is significantly more irregular than for the 1:1 case and,
consequently,
the spectral properties are closer to the GOE behavior. 

The application of an additional static electric field 
to the system has allowed us to enhance chaos in the secular motion.
Moreover,
the static electric field, for a generic orientation, breaks any
anti-unitary
symmetry of the system which has a dramatic effect on statistical 
properties of quasienergy levels. This is the first, to our knowledge,
experimentally realizable quantum system, with corresponding chaotic
classical
behavior, which exhibits breaking of any generalized time-reversal
symmetry.

We have studied the principal 1:1 resonance
for two slightly different static field orientations: the first one
corresponds to the anti-unitary invariance case, the other one to
breaking such a
symmetry. The classical phase space structures, in both cases, are
similar with 
predominately chaotic behavior. However, the statistical 
properties of the quantum spectrum change from a near-GOE to a 
near-GUE behavior when one switches from the preserved 
to broken anti-unitary 
symmetry case. In the intermediate situations, we could observe
a partial symmetry breaking effect due to the finite size of matrices
involved
in the problem.

We have studied the limit of small field amplitudes as it allows us to 
employ quantum perturbation theory. For higher amplitudes, 
the classical effective Hamiltonian
is known, but finding the quasienergy spectrum requires {\it full} 
quantum numerical calculations. For a pure microwave perturbation,
however, 
we may predict that 
statistical properties of the quantum spectrum, for stronger field,
should 
be the same as for weak field limit. This is because of the
specific form of the effective Hamiltonian, Eq.~(\ref{semi}), which 
depends on dynamical variables only through the $\Gamma_m$.

\section{Acknowledgements}

We are grateful to Felix~Izrailev for the permission to use his code 
for his NNS distribution. 
We thank the referee for bringing references
\cite{pandey83,mehta83,bohigas95} to our attention.
Support of KBN under project 2P302B-00915 (KS and JZ) is acknowledged. 
KS acknowledges support by the Alexander von Humboldt Foundation.
The additional support of the bilateral Polonium and PICS programs is
appreciated.
Laboratoire Kastler Brossel de
l'Universit\'e Pierre
et Marie Curie et de l'Ecole Normale Sup\'erieure is
UMR 8552 du CNRS.
\section{APPENDIX A} 
The amplitude and the phase of the atomic dipole at the microwave
frequency, $\Gamma_m(L,\Psi,M,\Phi;\alpha)$ and
$\delta(L,\Psi,M,\Phi;\alpha)$ appearing in Eq.~(\ref{EPr}) may be
expressed as \cite{sacha98c}
\bea
\Gamma_{m}&=&\Bigl\{
\left(\frac{1+\alpha}{2}\right)^{2}\left[V_{m}^2+U_m^2\right]
+\left(\frac{1-\alpha}{2}\right)^{2}\left[V_{-m}^2+U_{-m}^2\right]
\nonumber\\
&&+\frac{1-\alpha^2}{2}\Bigl[
\left(V_mV_{-m}-U_mU_{-m}\right)\cos2\Phi
-\left(V_mU_{-m}+U_mV_{-m}\right)\sin2\Phi\Bigr]
\Bigr\}^{1/2},
\label{gam}
\eea
and
\be
\tan \delta=\frac{(1-\alpha)(V_{-m}\sin\Phi+U_{-m}\cos\Phi)-
(1+\alpha)V_{m}\sin\Phi+U_{m}\cos\Phi)}
{(1-\alpha)(V_{-m}\cos\Phi-U_{-m}\sin\Phi)+
(1+\alpha)V_{m}\cos\Phi-U_{m}\sin\Phi)},
\ee
where $V_m$ and $U_m$ are Fourier expansion terms of
the original hamiltonian (\ref{h}) in action-angle variables.
Explicitly,  they read  
  \bea
 V_{0}(J,L,\Psi) &=& -\frac{3e}{2}J^{2}\cos\Psi,      \cr
 U_{0}(J,L,M,\Psi) &=& -\frac{3eM}{2L}J^{2}\sin\Psi,
\label{Va}
\eea
and for $m\ne 0$
\bea
 V_{m}(J,L,M,\Psi) &=& \frac{J^{2}}{m}[{\cal J}_{m}^{'}(me) +
 \frac{M\sqrt{1-e^2}}{Le}{\cal J}_{m}(me)]\cos\Psi,\cr
 U_{m}(J,L,M,\Psi) &=& \frac{J^{2}}{m}[\frac{M}{L}{\cal J}_{m}^{'}(me) +
 \frac{\sqrt{1-e^2}}{e}{\cal J}_{m}(me)]\sin\Psi.
\label{Vb}
\eea

In the above formulae $e=\sqrt{1-L^2/J^2}$ is
 an eccentricity of the electronic ellipse while 
${\cal J}_{m}(x)$ and ${\cal J}^{'}_{m}(x)$ denote the ordinary
 Bessel function and
its derivative, respectively.

\section{APPENDIX B} 

We give here explicit expressions for various level
spacing distributions which have been used in the present paper.

The Poisson distribution, corresponding to a system with classically 
integrable dynamics \cite{berry77}, reads 
\be
P(s)=\exp\left(-s\right).
\ee
For an ergodic classical behavior, the quantum spectrum of a generic 
system is conjectured \cite{bohigas84} to have  a nearest neighbor
spacing (NNS) distribution (for the unfolded spectrum) similar to that
of the 
random matrices of the same universality class. 
The resulting NNS distributions
are quite complicated  (see e.g. \cite{haake90,bohigas91}). However, a
good approximation is given by the so called Wigner surmise, obtained
for
matrices of rank 2. These are:
\be
P(s)=\frac{\pi}{2}s\exp\left(-\frac{\pi}{4}s^2\right)
\ee
for an anti-unitary invariant (GOE) system and 
\be
P(s)=\frac{32}{\pi^2}s^2\exp\left(-\frac{4}{\pi}s^2\right)
\ee
for broken anti-unitary invariance (GUE).

The phenomenological Brody distribution \cite{brody73} 
which interpolates between Poisson and GOE
distributions reads
\be
P(s)=C (\beta+1)s^\beta\exp\left(-C s^{\beta+1}\right)
\ee
with 
\be
C=\left[\Gamma\left(\frac{\beta+2}{\beta+1}\right)\right]^{\beta+1},
\ee
$\beta=0$ (resp. $\beta=1$) corresponds to the extreme case
of the Poisson (resp. GOE) 
statistics. 

Another attempt towards a simple distribution interpolating between
different
ensembles, is due to Izrailev \cite{izrailev90,casati91} and reads
\be
P(s)=As^\beta(1+sB\beta)^{f(\beta)}\exp\left[\frac{-\pi^2\beta
s^2}{16}-\frac{\pi}{4}
(2-\beta)s\right]
\ee
where\
\be
f(\beta)=\frac{2^\beta(1-\beta/2)}{\beta}-0.16874,
\ee
and $A$, $B$ are constants that  ensure 
\be
\int P(s)ds=1, 
\ee
and 
\be
\int sP(s)ds=1.
\ee
It is claimed to work reasonably well for all possible intermediate
situations.

While there exist several other propositions in the literature, we list
only the so called
Berry--Robnik statistics \cite{berry84}.
 It may be derived assuming an independent superposition of Poisson
spectrum and spectra corresponding to random matrix predictions. If one 
deals with Poisson and only one GOE spectrum, the Berry--Robnik
distribution
reads \cite{berry84}
\be
P(s)=\left[2q(1-q)+\frac{\pi}{2}q^3s\right]
\exp\left[(q-1)s-\frac{\pi}{4}q^2s^2\right]+(1-q)^2\exp[(q-1)s]
\mbox{erfc}\left(\frac{\sqrt{\pi}}{2}qs\right)
\ee 
where $0\le q\le 1$ is the relative weight of the GOE spectrum. 
Classically, $q$
corresponds to the relative volume of the chaotic part of phase space.
The similar Berry--Robnik distribution for a superposition of a Poisson
spectrum
and one GUE 
spectrum is \cite{robnik97}
\bea
P(s)=\left\{\left[\frac{32}{\pi^2}q^4s^2+\frac{8}{\pi}q^2(1-q)s+
(1-q)^2\right]\exp\left(-\frac{4}{\pi}q^2s^2\right) 
\right. \cr
\left.
+\left[2q(1-q)-(1-q)^2qs\right]
\mbox{erfc}\left(\frac{2}{\sqrt{\pi}}qs\right)\right\}\exp\left[(q-1)s\right].
\eea

Let us present also the expression for the $U(W)$ function which has
been 
employed in a fine-scale representation of the deviation of the
numerical 
level spacing distribution from the best fitting theoretical
distribution.
The following function \cite{prozen93}
\be
U(W)=\arccos\sqrt{1-W},
\ee
where $W$ is the value of the cumulative level spacing distribution
$\int{P(s) ds}$, ensures
that, over the whole range of $W$, i.e. from 0 to 1, the standard
deviation
of numerical data is uniform and equal to $\delta U=1/(\pi\sqrt{N})$, 
where $N$ is the total number of spacings.

For completeness, let us finally define the spectral rigidity
$\Delta_3(L)$
as an average over the spectral range used in analysis (i.e. over $x_0$)
of
\be
\Delta_3(x_0,L)=L^{-1}{\mathrm min}_{A,B} \int_{x_0}^{x_0+L}
 dx (N(x)-Ax-B)^2,
\ee
where $N(x)$ is the integrated level density (a staircase function).



\begin{table}
\caption{Fitted parameters for different distributions as defined in
the Appendix. The data collected correspond to the hydrogen atom 
driven by an elliptically
polarized microwave field and are shown in Fig.~\protect{\ref{three}}
and 
 Fig.~\protect{\ref{four}}. For each set of levels, we fit the nearest
 neighbor spacing distribution or the spectral rigidity $\Delta_3$ 
 and show here the value of the free parameter
 for the best fit.
 } 
\begin{tabular}{lcccc}
 & \multispan{2} \mbox{1:1~resonance} & \multispan{2}
\mbox{2:1~resonance} \\
& $n_0=50-59$ & $n_0=97-102$ & $n_0=50-59$ & $n_0=97-102$ \\
\hline
$q$ (Berry-Robnik): & 0.56 & 0.47 & 0.77 & 0.73 \\
$q$ ($\Delta_3$): & 0.55 & 0.43 & 0.76 & 0.73 \\
$\beta$ (Izrailev): & 0.17 & 0.12 & 0.42 & 0.37 \\
$\beta$ (Brody): & 0.23 & 0.16 & 0.49 & 0.44 \\
Time-reversal invariance & Yes & Yes & Yes & Yes \\
\end{tabular}
\label{t1}
\end{table}

\begin{table}
\caption{Fitted parameters for different distributions as defined in
the Appendix. The data  partially   shown in Fig.~\protect{\ref{seven}}
correspond to the hydrogen atom 
driven by an elliptically
polarized microwave field and exposed to an additional static
electric field. For each set of levels, we fit the nearest
 neighbor spacing distribution or the spectral rigidity $\Delta_3$ 
 and show here the value of the free parameter
 for the best fit.
 }
\begin{tabular}{lcccc}
 & \multispan{2} \mbox{$\varphi=0.4\pi$} & \multispan{2}
\mbox{$\varphi=\pi/2$} \\
& $n_0=50-59$ & $n_0=97-102$ & $n_0=50-59$ & $n_0=97-102$ \\
\hline
$q$ (Berry-Robnik): & 0.97  & 0.98   & 0.95 &  0.94 \\
$q$ ($\Delta_3$): & 0.98 & 0.98   & 0.96 & 0.95 \\
$\beta$ (Izrailev): & 1.73 & 1.83   & 0.85 & 0.82 \\
$\beta$ (Brody): & -- & -- & 0.85 & 0.83 \\
Time-reversal invariance & No & No & Yes & Yes \\
\end{tabular}
\label{t2}
\end{table}

\begin{table}
\caption{Fitted parameters for different distributions as defined in
the Appendix. The data  
correspond to the 2:1 resonance of the hydrogen atom 
driven by an elliptically
polarized microwave field and exposed to an additional static
electric field. For each set of levels, we fit the nearest
 neighbor spacing distribution or the spectral rigidity $\Delta_3$ 
 and show here the value of the free parameter
 for the best fit. The data with scaled energy  ${\cal H}\in(8.5,9.5)$
 and for ${\cal F}=5000$ are analyzed.
 }
\begin{tabular}{lcc}
& $n_0=99-101$ & $n_0=97-102$ \\
& \mbox{$\varphi=0.2,0.25,0.3\pi$} &
\mbox{$\varphi=\pi/2$} \\
\hline
$q$ (Berry-Robnik): & 0.94    &  0.94 \\
$q$ ($\Delta_3$): & 0.94   & 0.95 \\
$\beta$ (Izrailev): & 1.47   & 0.82 \\
$\beta$ (Brody): & --  & 0.83 \\
Time-reversal invariance  & No & Yes  \\
\end{tabular}
\label{t3}
\end{table}

\begin{figure}
\caption{Poincare surfaces of sections (at $\Phi=0$) of the classical
secular 
motion, Eq.~(\protect{\ref{purem}}), for the hydrogen atom perturbed by
a weak resonant microwave field with elliptical polarization
(polarization
parameter $\alpha=0.4$).
 The coordinates used for the plot are the scaled
total
angular momentum $L_0=L/n_0$ and its canonically conjugate angle $\Psi.$
Left column -- the 1:1 resonance case (microwave frequency equal to the
unperturbed Kepler frequency of the electron on its elliptical
trajectory) for
the secular energies (going from bottom to top), 
${\cal H}_{\mathrm eff}/F_0^2=$0.02, 0.03, 0.04, 0.07. Right column --
the 2:1 
resonance case (microwave frequency is twice the Kepler frequency), 
for ${\cal H}_{\mathrm eff}/F_0^2=$0.0006, 0.0013, 0.0018, 0.0021 again
from 
bottom 
to top. Note that, for the parameters chosen, not the whole $(L_0,\Psi)$
space is accessible.
}
\label{one}
\end{figure}

\begin{figure}
\caption{Energy levels of the $n_0=20$ hydrogenic
manifold resonantly driven by an elliptically polarized  
microwave field vs. the degree of polarization, $\alpha$, for the 1:1
resonance
case [panel (a)] and 2:1 resonance case [panel (b)]. We plot the 
shifts (divided by $F_0^2$) of the energy levels with respect to 
the unperturbed energy of the atom. The upper states evolve smoothly
with almost exact level crossings, in agreement with the classical
mostly regular dynamics. In contrast, most of the lowest states evolve
irregularly
and sometimes display large avoided crossings,
 a signature of a classically
chaotic behaviour.
}
\label{two}
\end{figure}

\begin{figure}
\caption{A part of the spectrum of $n_0=50$ hydrogenic
manifold resonantly driven by an elliptically polarized  
microwave field vs. the degree of polarization, $\alpha$, for the 2:1
resonance
case in the region of mixed regular and chaotic motion. We plot the 
shifts (divided by $F_0^2$) of the energy levels with respect to 
the unperturbed energy of the atom. Most of the states display large 
avoided crossings, a signature of a classically
chaotic behaviour.
}
\label{twobis}
\end{figure}

\begin{figure}
\caption{ Cumulative level spacing distribution, $W(s)$, for the pure
microwave 
perturbation (degree of the polarization $\alpha=0.4$)
for the 1:1 resonance case. Levels in the range of
 ${\cal H}_{\mathrm eff}/F_0^2=0.02-0.045$ are analyzed. 
In panel (a), the 
solid line represents numerical data for $n_0=50-59$
(there are about 10000 spacings); the
dashed and dash-dotted lines represent the Poisson and GOE
distributions 
respectively. Panel (b) shows a fine-scale representation of 
the deviation of the numerical level spacing distribution 
from the best 
Izrailev distribution in terms of the $U(W)-U(W_{\mathrm Izrailev})$
vs. $W$; the transformation $U(W)=\arccos\sqrt{1-W}$ is used in order to 
have uniform statistical error over the plot - compare
{\protect\cite{prozen93}}
and the Appendix. 
The upper
and lower noisy curves represent one standard deviation from the
calculated data
which lie in the middle of the band. The long-dashed curve represents
the best
Berry-Robnik
distribution while the dotted line the best Brody distribution. 
Panel (c) , the 
solid line represents the numerical data for $n_0=97-102$
(there are about 20000 spacings),
the
dashed and dash-dotted lines represent the Poisson and GOE
distributions as in (a). 
Panel (d) shows the corresponding 
deviations of the numerical level spacing distribution from the best
fits  
-- the notation is the same as in panel (b).
Note that the best fitted Berry-Robnik or Izrailev distributions
are altogether in excellent agreement with the numerical results. At the
scale
of (a) and (c), they are not distinguishable from the data.
}
\label{three}
\end{figure}
 
\begin{figure}
\caption{ The same as in Fig.~\protect{\ref{three}}, but for the 2:1
resonance
case for the range of 
${\cal H}_{\mathrm eff}/F_0^2=0.00035-0.0018$. 
}
\label{four}
\end{figure}

\begin{figure}
\caption{ Spectral rigidity $\Delta_3$ for the pure microwave 
perturbation ($\alpha=0.4$) compared with the random
ensemble predictions. Numerical data (solid
lines) for $n_0=50-59$ are shown in panel (a) for the 1:1 resonance case
and 
in panel (b) for the 2:1 resonance. 
The data for $n_0=97-102$ are presented in panels 
(c) and (d) for the 1:1 and 2:1 resonances respectively. The dotted
lines 
are the fits of Eq.~(\protect{\ref{delta3}}) while the dashed lines
indicate
the Poisson (upper straight lines) and GOE (lower curves) predictions. 
Note the saturation at large $L$, in agreement with the semiclassical
prediction. }
\label{five}
\end{figure}

\begin{figure}
\caption{Poincar\'e surfaces of section (at $\Phi=0$) of the classical
secular 
motion, Eq.~(\protect{\ref{finscal}}), for the hydrogen atom in a static
electric field
and driven by an elliptically polarized ($\alpha=0.4$)
microwave field resonant with the Kepler
frequency (1:1 resonance). 
Left and right columns correspond to a static electric field
with scaled amplitude ${\cal F}=10$ and
different orientations $\varphi=0$, $\theta=\pi/4$ and 
$\varphi=\pi/2$, $\theta=\pi/4$ respectively.
The secular energies (going from bottom to top, the same for left and
right panels) are:
${\cal H}=-0.2,\ 0.4,\ 0.8,\ 1.4$. 
Note that, for the parameters
chosen, not the whole $(L_0,\Psi)$
space is accessible.}
\label{six}
\end{figure}

\begin{figure}
\caption{ Cumulative level spacing distribution, $W(s)$, for the the
hydrogen
atom in an elliptically polarized microwave field
(ellipticity parameter $\alpha=0.4$) combined with a static electric
field, 
for the 1:1 resonance case. Levels in the range of ${\cal H}=0-1.4$ and
for
$n_0=50-59$ are analyzed (about 20 000 spacings). The value of the 
reduced microwave strength 
is chosen as ${\cal F}=10,$ while the angle is $\theta=\pi/4$. 
Panel (a) shows the data
(solid line) for $\varphi=0.4\pi$, i.e. for broken generalized time
reversal
invariance. The fitted Izrailev distribution cannot be distinguished
from
numerical data. Dashed and dash-dotted lines correspond to GUE and GOE
predictions respectively
(thus the data are closer to GUE than GOE). Panel (b) shows the
 difference between the numerical result and the best Izrailev
distribution 
(horizontal line). 
The upper and lower noisy curves yield one standard
deviation from  the numerical data that lie in the middle of the band.
The Izrailev distribution stays well within the one standard
deviation practically everywhere. 
The dashed line corresponds to a Berry-Robnik distribution
(for the mixture of Poisson and GUE spectra) which is clearly inferior
to
the Izrailev distribution. Panel (c) and (d) correspond to
$\varphi=\pi/2$ where
the anti-unitary symmetry is restored. In (c), dashed line corresponds
now to
Poisson, the data trace the distribution close to GOE one. In panel (d)
the
dashed-dotted line corresponds to Brody distribution. Clearly Brody and
Izrailev
distributions practically coincide and work much better than the 
Berry-Robnik
distribution.     
}
\label{seven}
\end{figure}

\begin{figure}
\caption{Spectral rigidity, $\Delta_3$  compared with the random
ensemble predictions for the hydrogen atom resonantly driven by an
elliptically
polarized microwave field in the presence of a static electric field
of different
orientation. The numerical data used are the same as in
figure~\protect{\ref{seven}}. 
Panel (a) corresponds to broken anti-unitary symmetry
($\varphi=0.4\pi$), 
panel (b) to $\varphi=\pi/2$ (generalized time-reversal invariant case).
The dotted lines 
denote the fit of Eq.~(\protect{\ref{delta3}}) for (b) and a similar
expression for (a) with GUE in place of GOE. The dashed lines indicate
the Poisson (upper straight lines), GOE and GUE predictions. Observe
that, for
broken  anti-unitary  
symmetry, the data trace between the GOE and GUE
curves. 
The typical saturation effects appear for large $L$.
  }
\label{eight}
\end{figure}

\begin{figure}
\caption{The  gradual symmetry
breaking with the orientation of the electric field, i.e. $\varphi$.
Filled circles correspond to the fitted Izrailev repulsion parameters
$\beta$
for data around $n_0=55$, open circles for data around $n_0=100$. 
Close to $\varphi=0.5\pi$ -- where an anti-unitary symmetry exists --
there is sudden drop in $\beta$ by roughly one unit, as predicted
by Random Matrix Theory. For smaller (larger -- the figure is symmetric
around
$\pi/2$) 
values of $\varphi$, the classical motion is less chaotic with
large regular islands in the phase space -- this explains the decrease
of
 $\beta$ for $\varphi <0.4\pi (>0.6\pi)$.
   For a further discussion, see text.
}
\label{nine}
\end{figure}

\begin{figure}
\caption{ Nearest neighbor spacing distribution for the hydrogen atom
resonantly driven by an elliptically polarized  microwave field 
in the presence of a static field oriented in the direction
($\varphi=0.4\pi$, $\theta=\pi/4$), such that the generalized
time-reversal
symmetry is broken. The numerical data correspond to $n_0=97-102$
(40 000 spacings), the solid line
gives the fitted Izrailev distribution. The dash-dotted and dashed lines
correspond to GOE and GUE Wigner surmises, respectively. The numerical
results
clearly show a stronger level repulsion than in the GOE case, a
signature
of the breaking of any anti-unitary symmetry.
}
\label{ten}
\end{figure}


\begin{thebibliography}{10}
\bibitem{remark1}
For example, the static electric field used in Fig.~10(a)-(d) of
reference
\protect{\cite{main98}} or in Fig.~13 of \protect{\cite{uzer97}} in
order
to produce a significant chaotic component are so high that the
corresponding
bound states do not exist any longer, but turn into very broad
unresolved
resonances.
\bibitem{remark2}
Equivalently, one can define in the Hilbert space a new scalar product
for which the Sturmian basis is orthonormal. The $1/2r$ factor disappears
in Eq.~(\protect{\ref{unity}}), but all operators have to be multiplied
by $2r$ to preserve hermiticity. The choice of one or the other
formulation is just a matter of convenience, they are totally
equivalent.
\end{thebibliography}
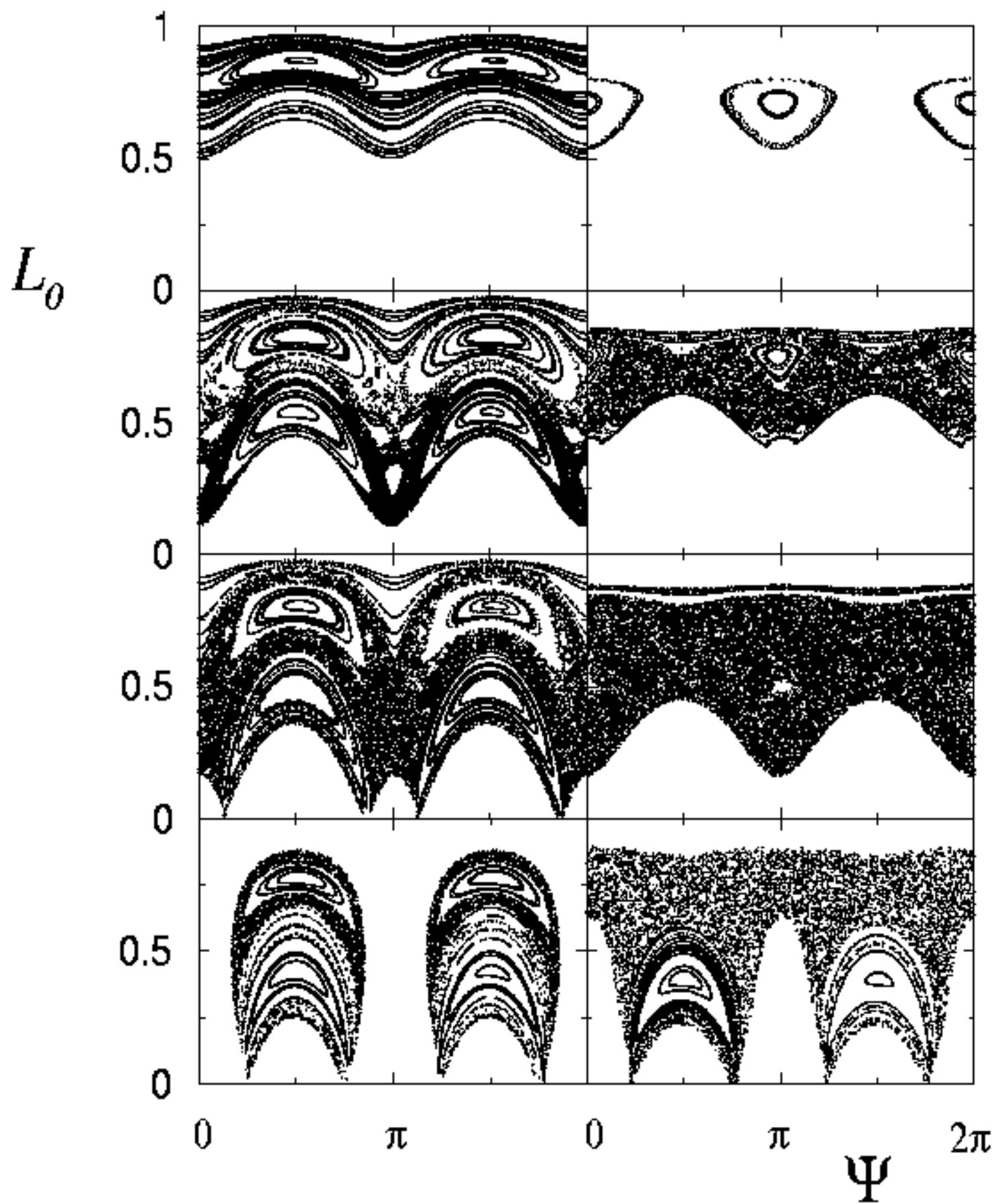
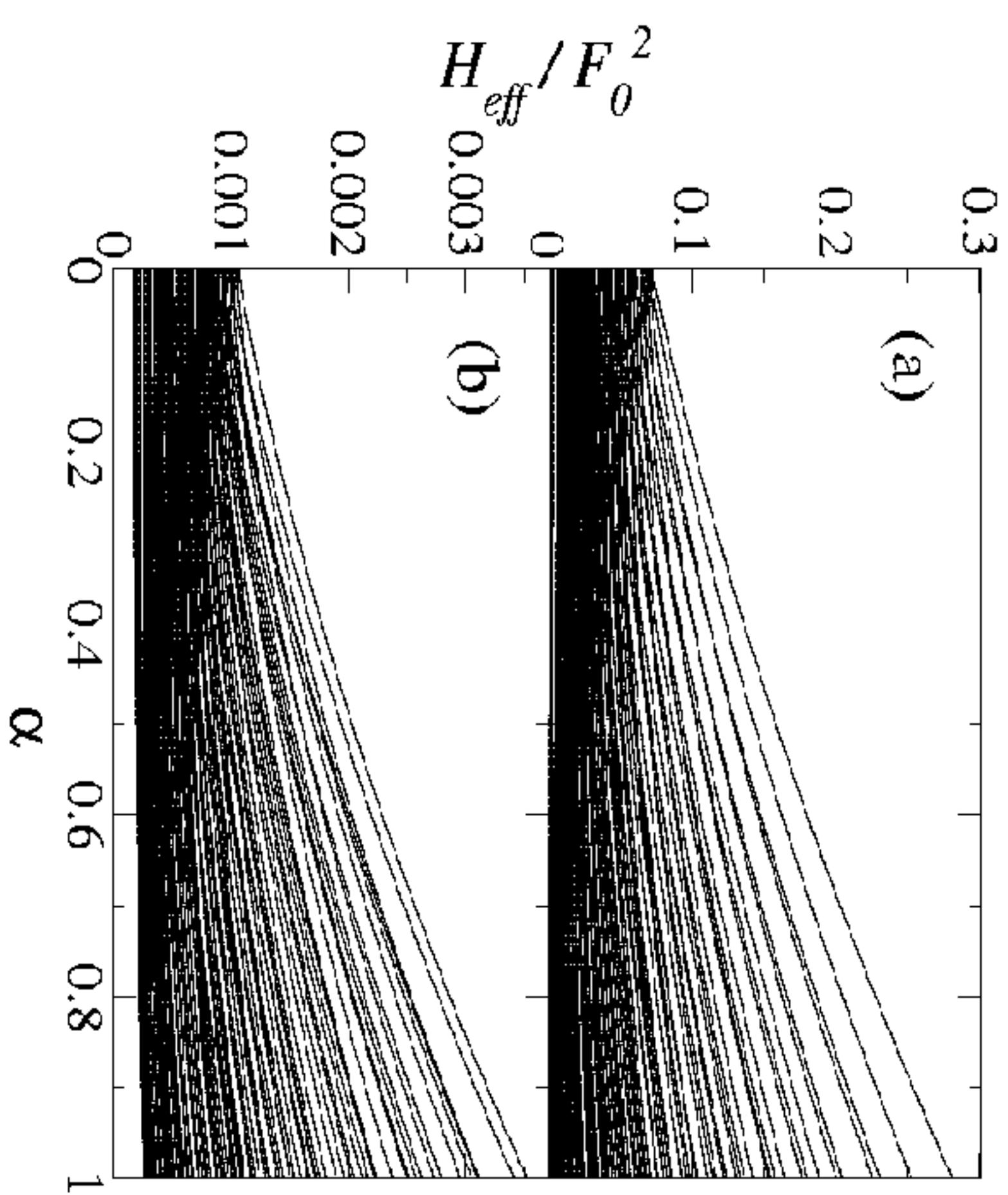
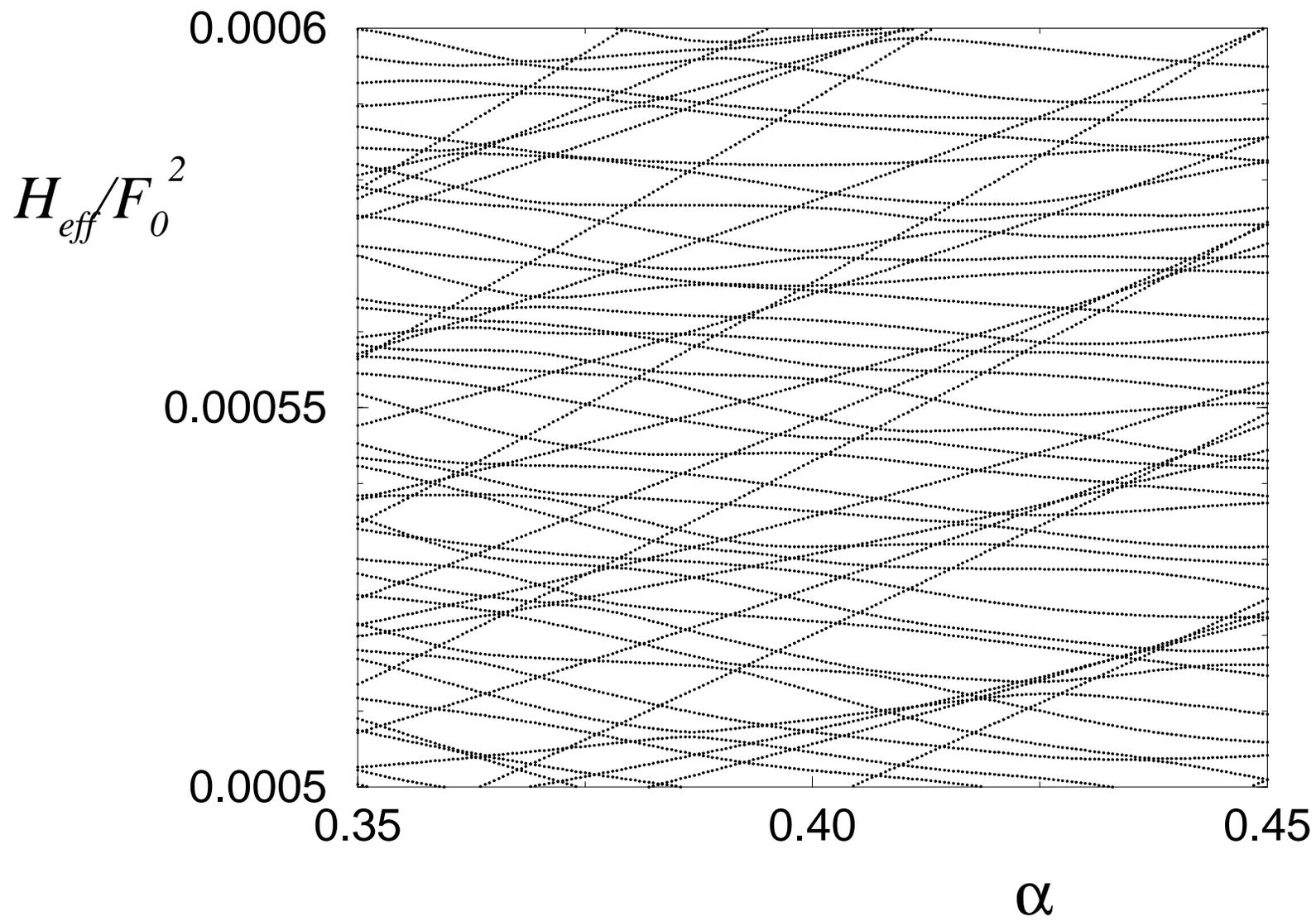
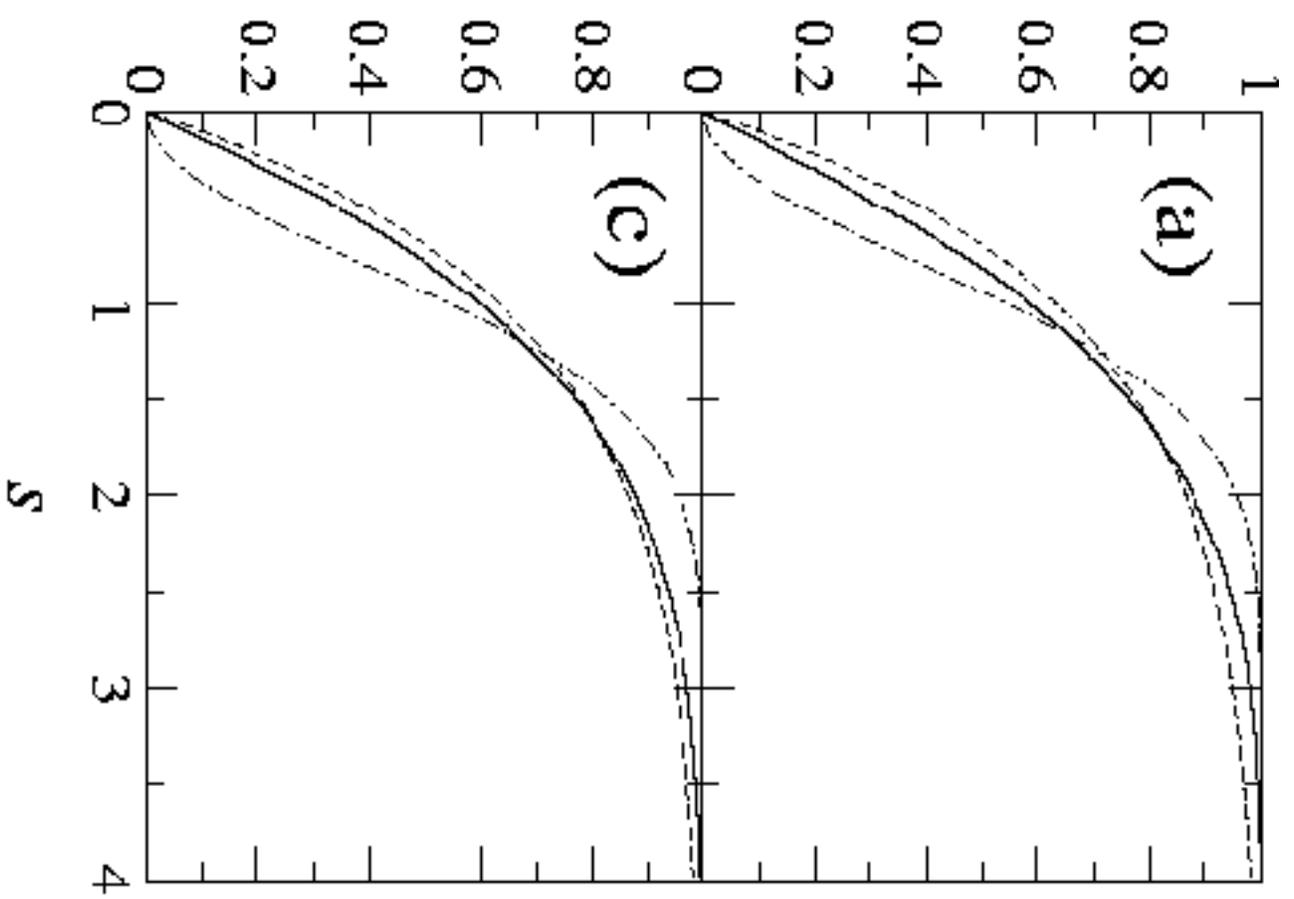
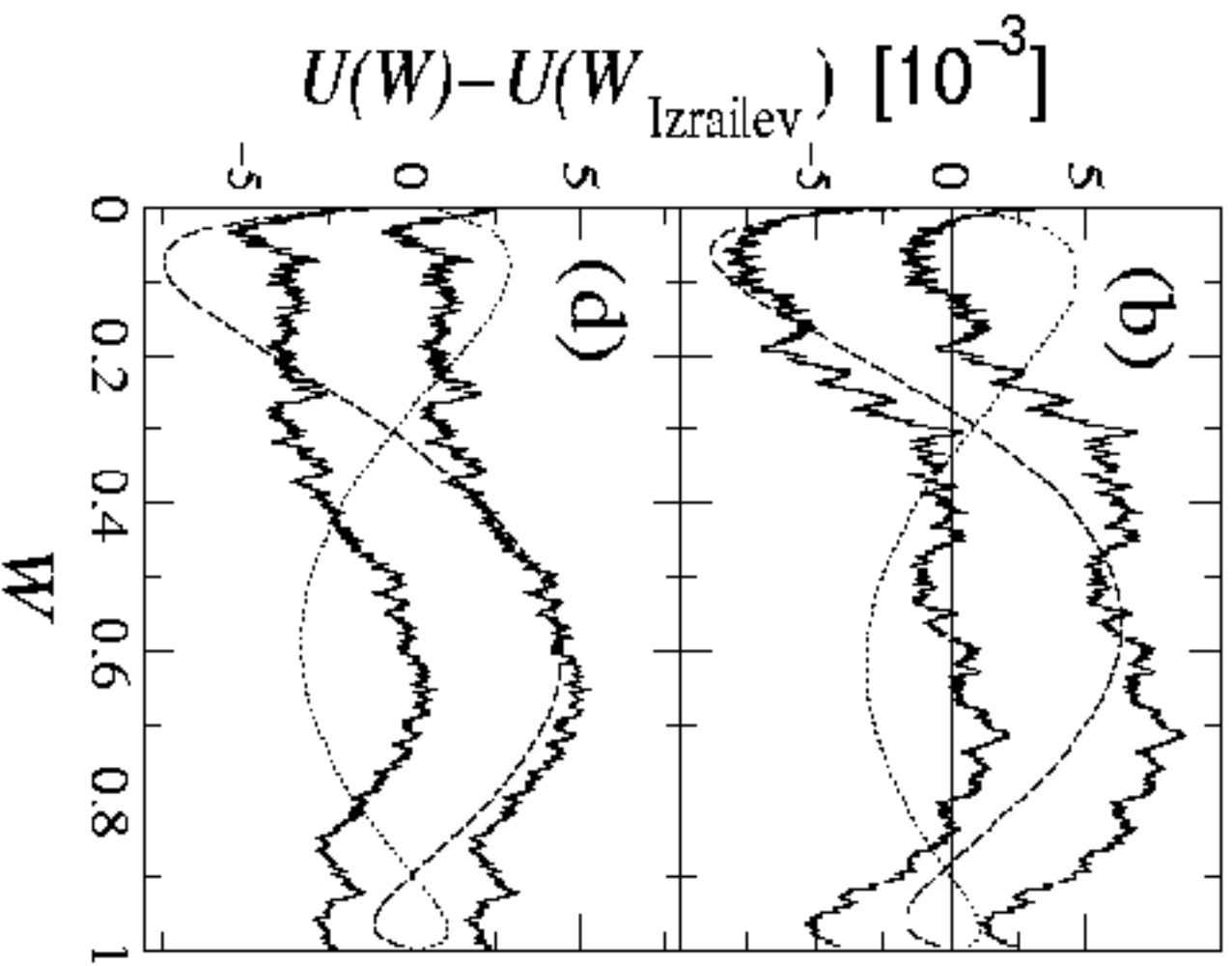
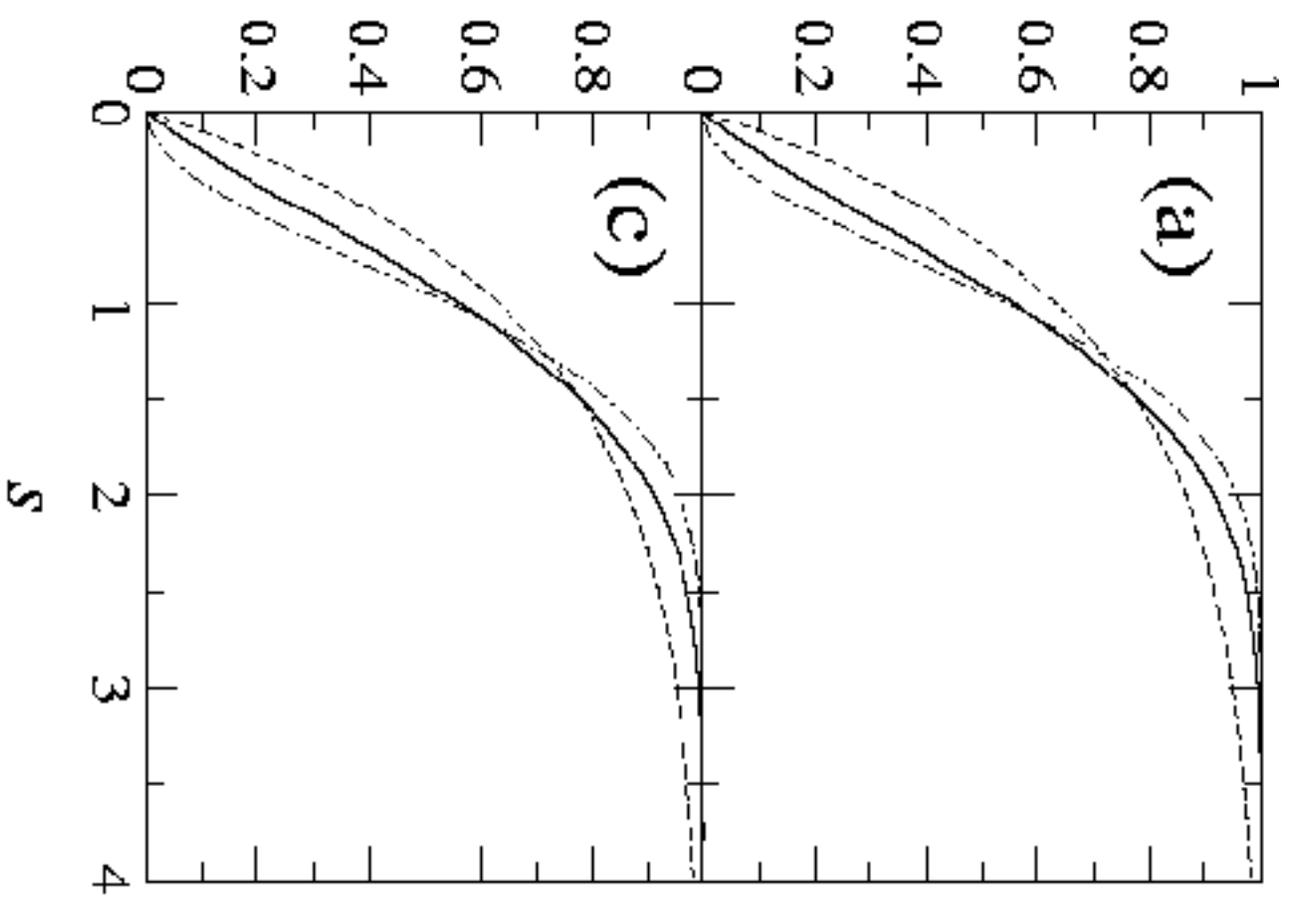
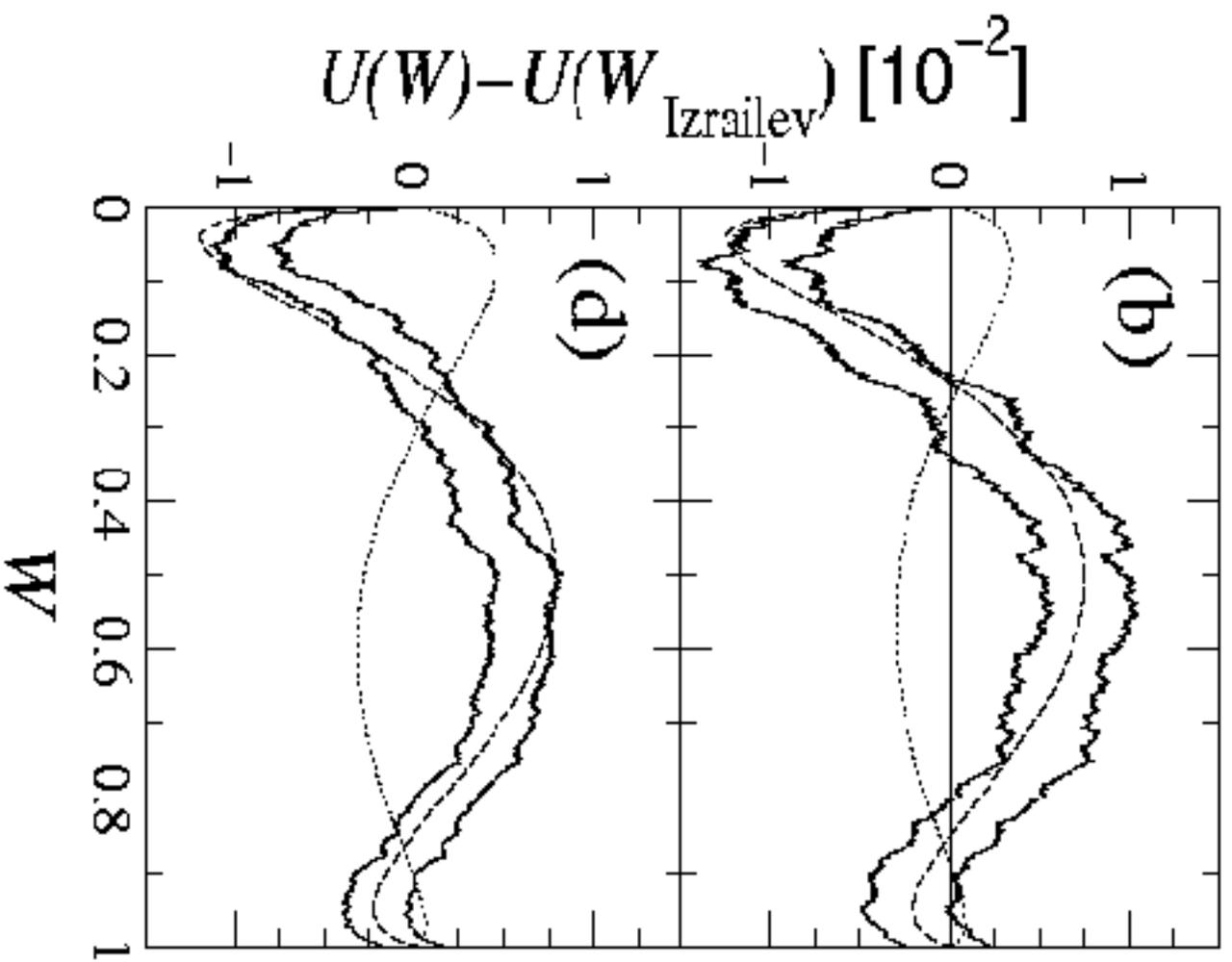
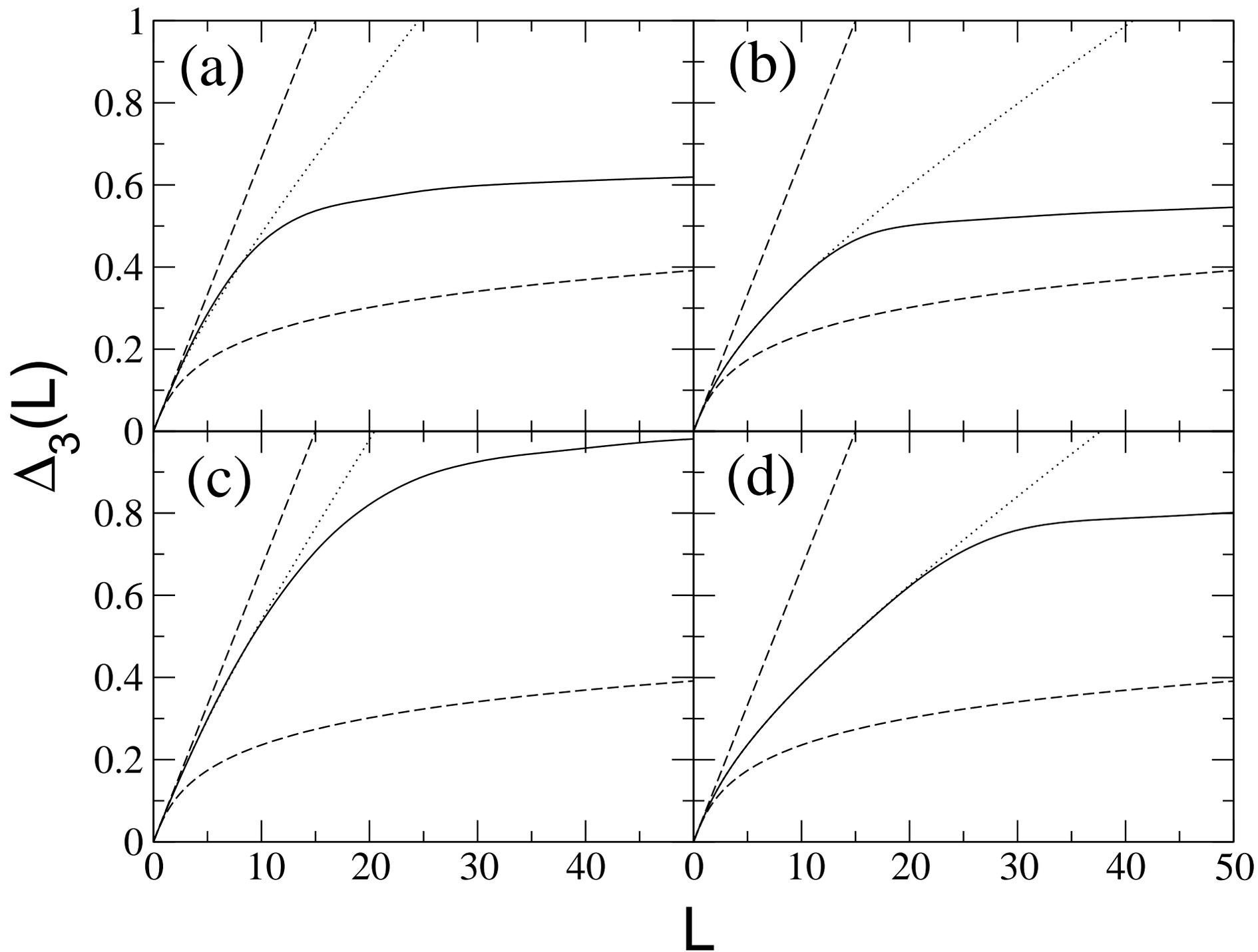
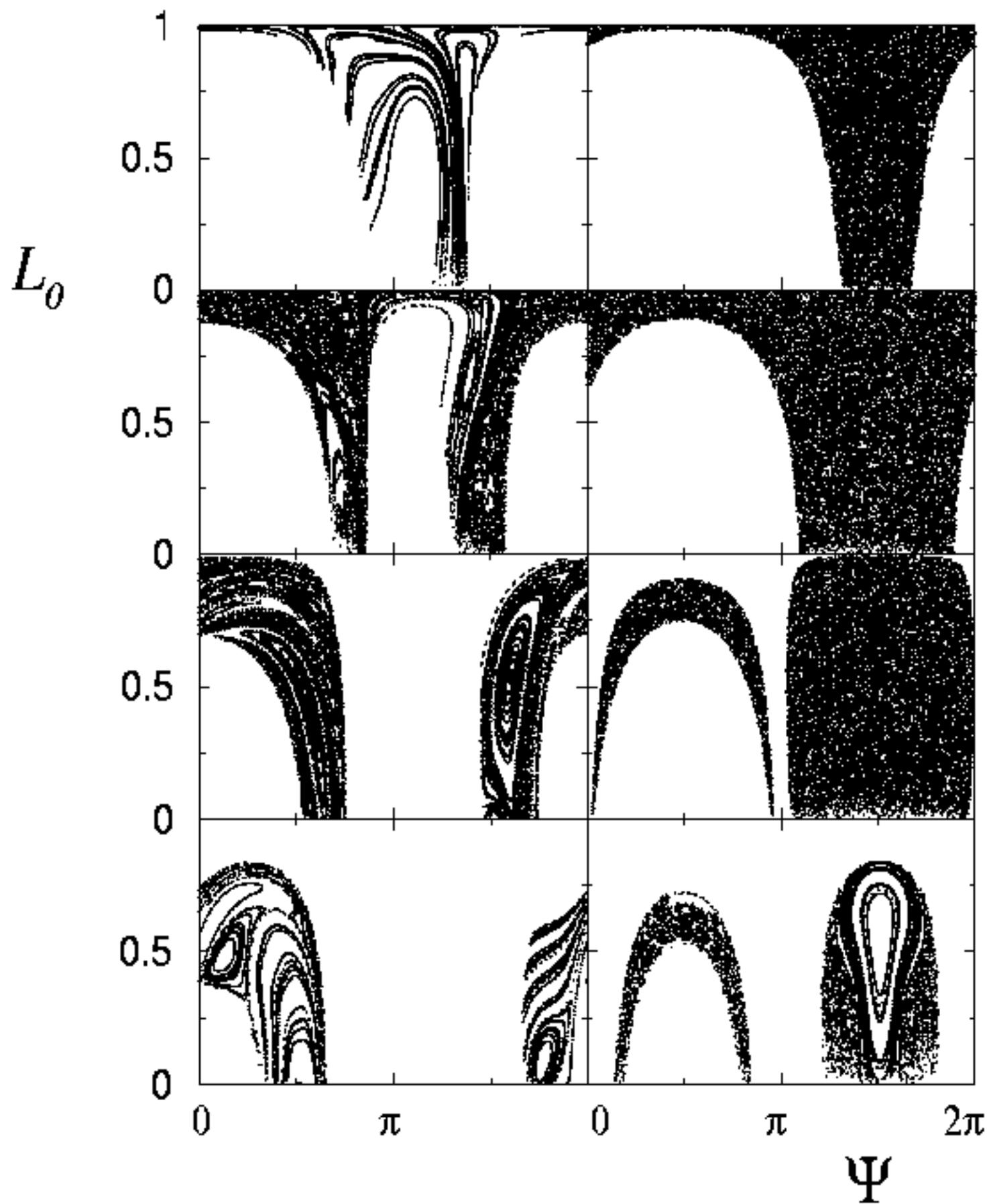
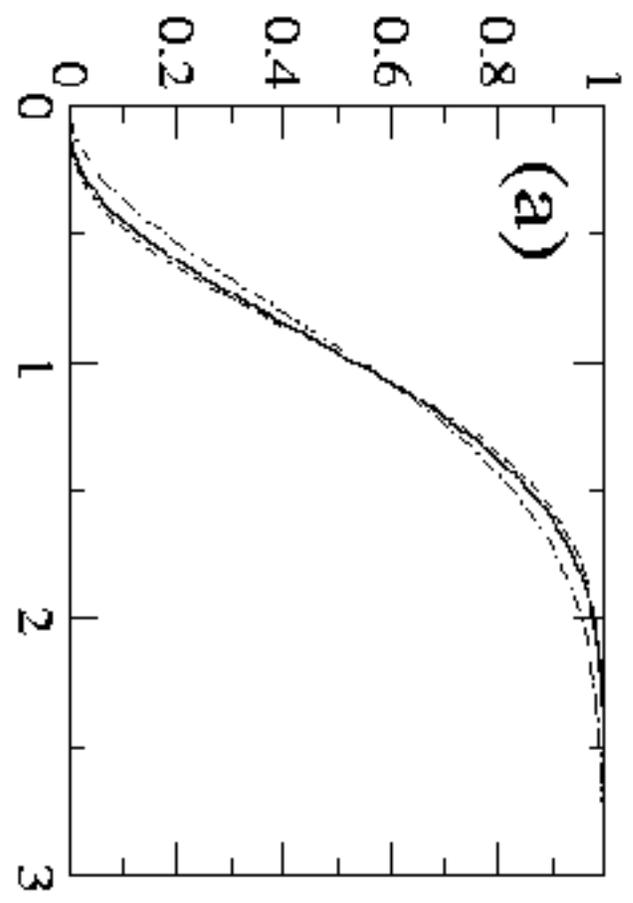
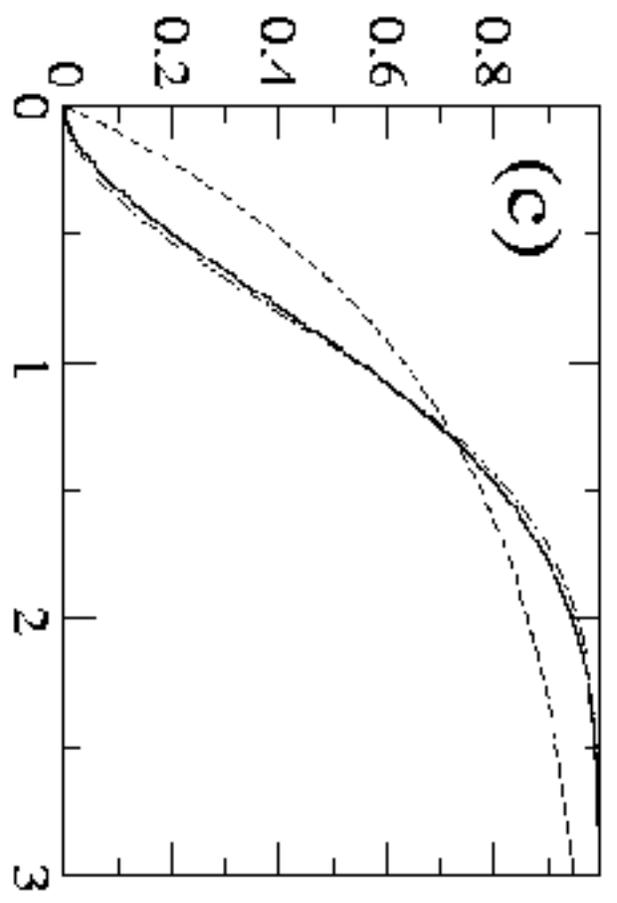
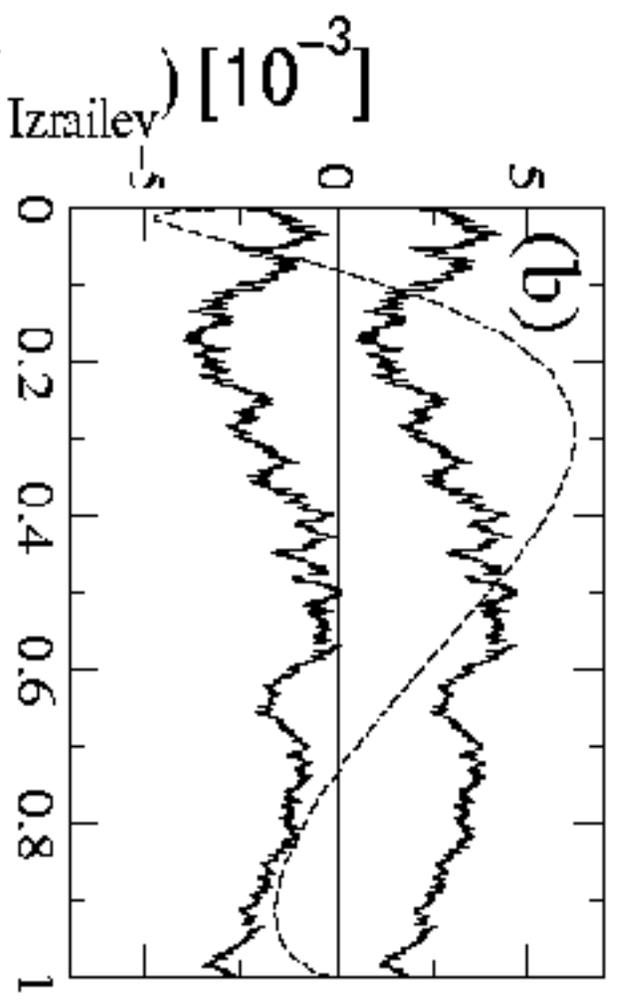
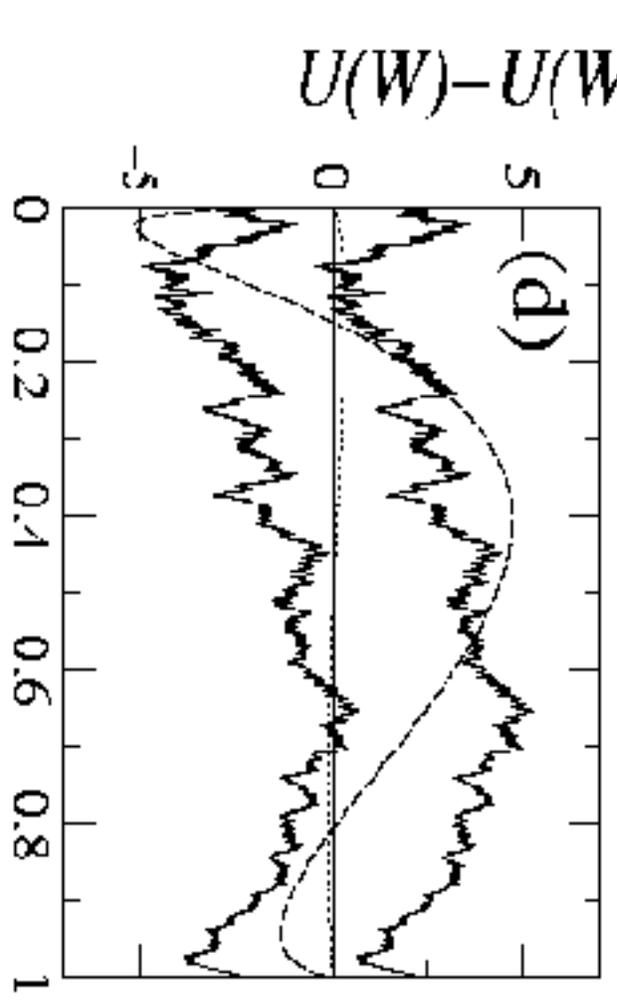
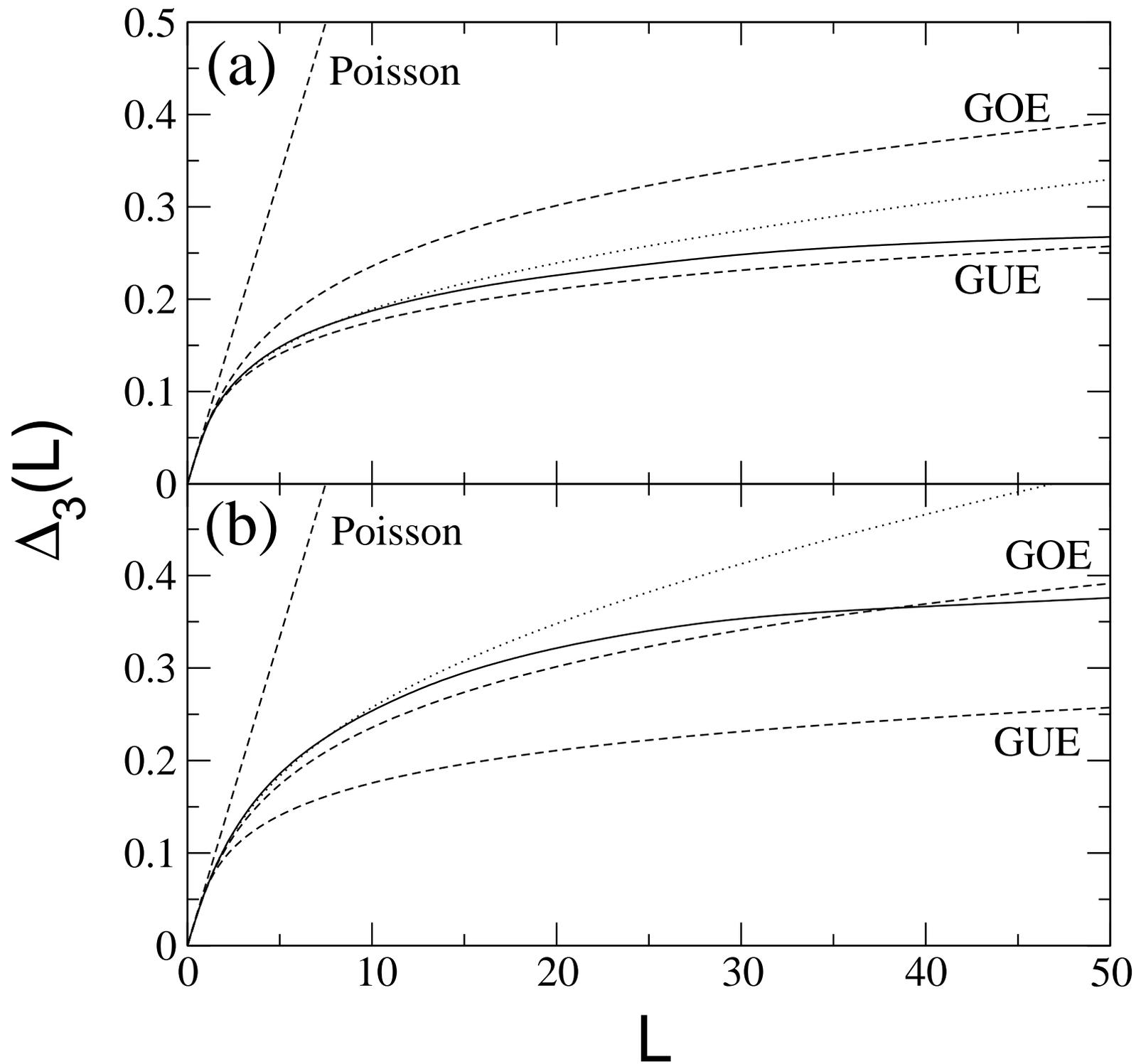
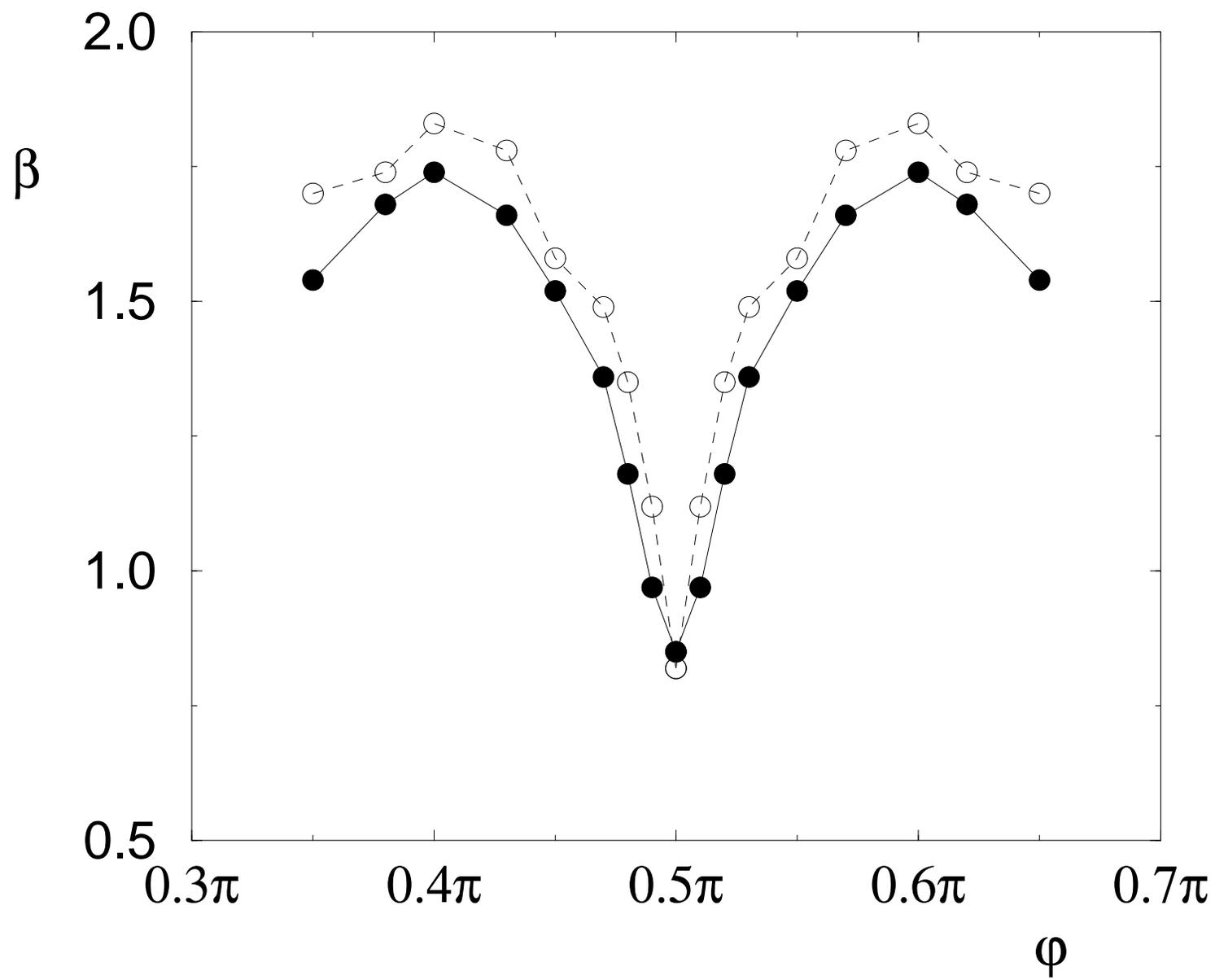
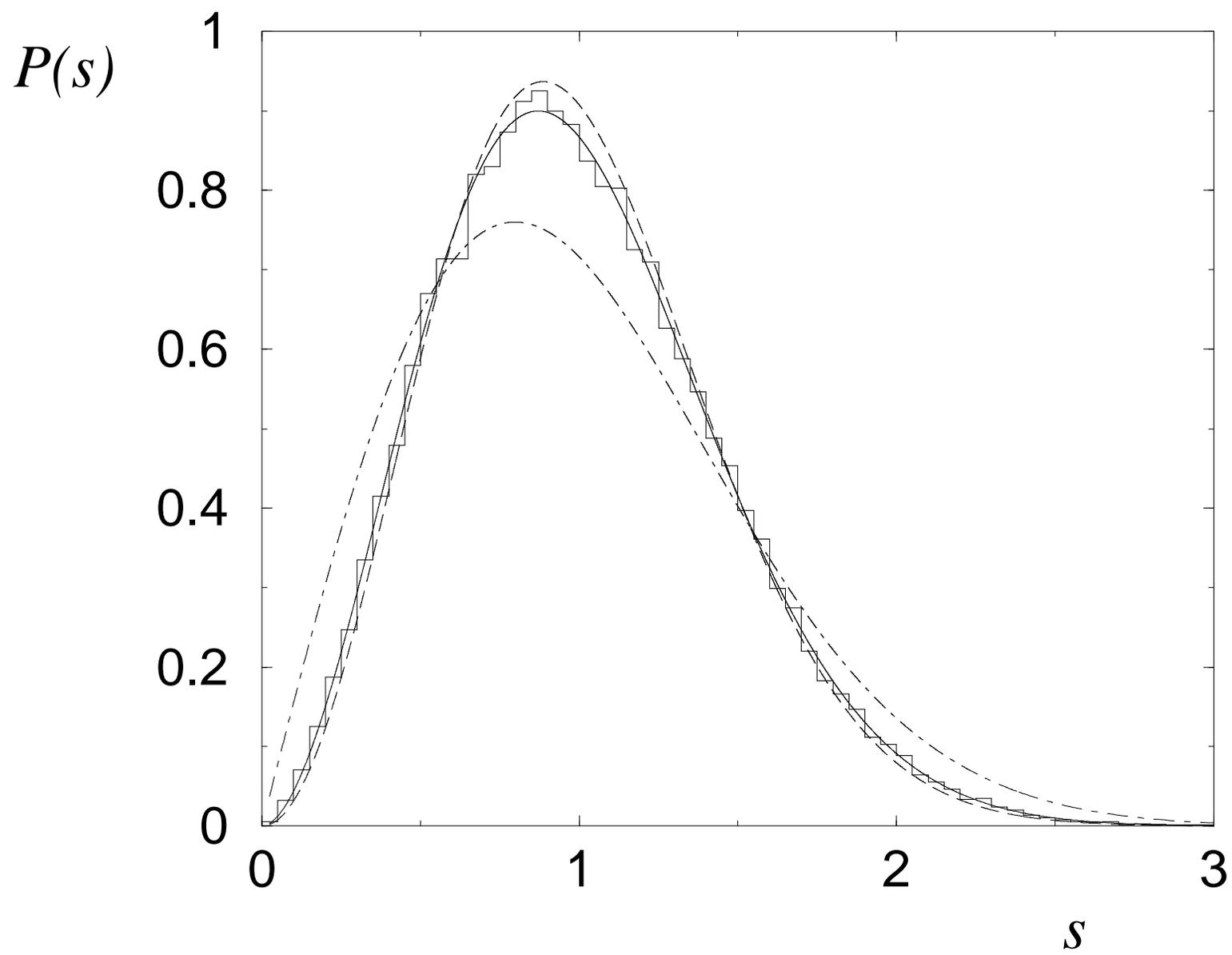
\end{document}